\documentclass[twocolappendix]{emulateapj}
\usepackage{graphicx}
\usepackage{amsmath}
\usepackage{gensymb}

\newcommand{\ip}{$i_{\text{P1}}$}
\newcommand{\rp}{$r_{\text{P1}}$}
\newcommand{\Iacs}{$I_{\text{ACS}}$}
\newcommand{\Iacsip}{$I_{\text{ACS-PS1}}$}

\newcommand{\av}{$A_V$}

\providecommand{\given}{\ensuremath{\,\mid\,}}
\providecommand{\prob}{\ensuremath{\mathcal{P}}}
\providecommand{\set}[1]{\ensuremath{\left\{\,#1\,\right\}}}

\begin{document}

\title{Panchromatic Hubble Andromeda Treasury. XIV. The Period-Age Relationship of Cepheid Variables in M31 Star Clusters}
\author{Peter Senchyna\altaffilmark{1}, L. Clifton Johnson\altaffilmark{1}, Julianne J. Dalcanton\altaffilmark{1}, Lori C. Beerman\altaffilmark{1}, Morgan Fouesneau\altaffilmark{2}, Andrew Dolphin\altaffilmark{3}, Benjamin F. Williams\altaffilmark{1}, Philip Rosenfield\altaffilmark{4}, S{\o}ren S. Larsen\altaffilmark{5}}

\email{senchp@u.washington.edu}
\altaffiltext{1}{Department of Astronomy, University of Washington, Box 351580, Seattle, WA 98195, USA}
\altaffiltext{2}{Max Planck Institute for Astronomy, Koenigstuhl 17, D-69117 Heidelberg, Germany}
\altaffiltext{3}{Raytheon Company, 1151 East Hermans Road, Tucson, AZ 85756, USA}
\altaffiltext{4}{Department of Physics and Astronomy G. Galilei, University of Padova, Vicolo dell'Osservatorio 3, I-35122 Padova, Italy}
\altaffiltext{5}{Department of Astrophysics/IMAPP, Radboud University P.O. Box 9010, 6500 GL Nijmegen, The Netherlands}

\shorttitle{PHAT/PS Cluster Cepheids}
\shortauthors{Senchyna et al.}

\begin{abstract}
We present a sample of 11 M31 Cepheids in stellar clusters, derived from the overlap of the Panchromatic Hubble Andromeda Treasury (PHAT) cluster catalog and the Pan-STARRS1 (PS1) disk Cepheid catalog.
After identifying the PS1 Cepheids in the HST catalog, we calibrate the PS1 mean magnitudes using the higher resolution HST photometry, revealing up to 1 magnitude offsets due to crowding effects in the ground-based catalog.
We measure ages of the clusters by performing single stellar population fits to their color-magnitude diagrams (CMDs) excluding their Cepheids.
From these cluster age measurements, we derive an empirical period-age relation which agrees well with the existing literature values.
By confirming this relation for M31 Cepheids, we justify its application in high-precision pointwise age estimation across M31.
\end{abstract}


\newpage
\section{Introduction} \label{intro}

Cepheids are young pulsating variable stars, widely-known for their utility in distance measurement.
They exhibit a tight period-luminosity (PL) relation \citep[the Leavitt Law,][]{Leavitt1912}, a product of the fact that this form of pulsation only occurs for stars in an unstable evolutionary phase occupying a thin strip in luminosity-effective temperature space.
Cepheids also exhibit a period-age relation, as more massive stars will evolve through the Cepheid strip earlier and at brighter magnitudes \citep{Kippenhahn1969}.
Thus, Cepheids can act as standard candles as well as useful probes of star formation history in the local universe.

Testing the period-age relation directly requires an age estimate independent of the Cepheids.
These ages are generally obtained through color-magnitude diagram (CMD) fitting applied to a single-age stellar population (SSP), which limits rigorous testing to well-resolved stars with known distances that are close enough for the main sequence turnoff to be visible.
Thus, previous tests of period-age relations have focused on Cepheid samples in our Galaxy or in the Small and Large Magellanic Clouds \citep[MCs: SMC and LMC, respectively; see for instance,][]{Bono2005,Efremov2003}.
The MCs are well-resolved in ground-based surveys and have large samples of Cepheids and stellar clusters, making them well-suited for CMD age estimation.
However, they have significantly lower metallicity than the more massive galaxies in the Local Group where Cepheid distance and age measurements are often applied.

The Andromeda galaxy (M31) provides an excellent opportunity to constrain the period-age relation at high metallicity.
This spiral galaxy is close enough that individual bright stars in it are well-resolved by the Hubble Space Telescope (HST), even within stellar cluster environments.
Existing period-age relations have been used to infer disk star formation in M31 \citep[for instance,][]{Magnier1997}, but not yet directly calibrated there.
Thanks to dedicated CCD surveys, large samples of M31 Cepheids have been obtained in recent years.
The DIRECT project, summarized by \citet{Macri2004}, searched for Cepheids and detached eclipsing binaries in a number of M31 fields.
\citet{Vilardell2007} identified a sample of 416 Cepheids in the process of searching for eclipsing binaries, with photometry sufficient to characterize the pulsation modes of 356 Cepheids.
The largest M31 Cepheid sample to-date was obtained by \citet{Kodric2013}, using the Pan-STARRS 1 (PS1) survey telescope PAndromeda dataset \citep{Lee2012}.
This sample avoids many of the issues which plagued earlier searches, namely: limited spatial coverage, few epochs, and short observational campaigns (which can introduce bias against finding Cepheids in certain period ranges).
The PS1 Cepheids were found via single-camera differential photometry covering the entire M31 disk.
The photometry were obtained in 183 epochs, spanning a year.
The lightcurves are well-sampled, and allow for robust period determination and Cepheid characterization.
A total of 1440 of these Cepheids were identified as fundamental-mode pulsators.

Progress has also been made on individual-star photometry in M31.
The Panchromatic Hubble Andromeda Treasury, or PHAT \citep{Dalcanton2012,Williams2014}, is a large survey of a third of the M31 disk in 6 filters with HST.
This survey provides another means to study individual Cepheids found using high-cadence ground-based photometry.
For instance, \citet{Riess2012}, \citet{Kodric2015}, and \citet{WagnerKaiser2015} have used the PHAT dataset to constrain the NIR period-luminosity relationship and improve the determination of the M31 distance modulus.
PHAT observations have also enabled the development of a large and well-characterized star cluster sample 
\citep{Johnson2015}, most of which are well-resolved in the HST data.

In this paper, we measure ages for individual Cepheids by identifying them in stellar clusters with secure measured ages.
Specifically, we cross-correlate the \citet{Kodric2013} Cepheid catalog with the \citet{Johnson2015} star cluster catalog.
The final sample of 11 high-quality candidates adds to the approximately 23 Galactic open cluster Cepheids \citep{Anderson2013}, together providing a solar metallicity complement to samples in the LMC and SMC \citep[e.g.,][]{Pietrzynski1999}.

In Section \ref{data}, we describe the PHAT and PS1 data products which we utilize.
Section \ref{analysis} describes the identification of the Cepheids in the PHAT data, the process and results of dephasing the PS1 magnitudes and the determination of isochrone fits for the clusters.
In Section \ref{discussion} we compare these cluster age determinations with several period-age relations in the literature.
We conclude with a brief summary in Section~\ref{conclusion}.

\section{Observational Data} \label{data}

\subsection{PAndromeda Cepheids}

We select Cepheids from the Pan-STARRS Cepheid sample presented by \citet{Kodric2013}, which is large and well-characterized.
Observations were made in the \ip{} and \rp{} filters over 183 epochs from 2010 July 23 to 2011 August 12 (with a half-year gap in the schedule).
Before inclusion in the final sample, the candidates passed cuts based on CMD location and lightcurve Fourier parameters.
\citeauthor{Kodric2013} manually classified a subset of the lightcurves and defined regions in Fourier space corresponding to fundamental mode, first-overtone, and Type-II Cepheids based on these classifications.
This enabled them to automatically classify all of the objects in their catalog.

The PS1 lightcurves are well-sampled, allowing us to extend them to the epoch of HST observation with minimal error.
We can thus correct for the phase of the Cepheid at the times of our HST observations, and derive accurate mean magnitudes from even the small number of HST observations. 
The process by which we extend the PS1 lightcurves is described in Section~\ref{analysis}.

\subsection{PHAT Clusters}

The most reliable way to assign ages to Cepheids is to associate them with a single age stellar population.
At the distance of M31, the main sequence turnoff for even young populations is near the HST detection limit.
Though it is possible to assign ages to recent bursts of star formation in any region of the M31 disk \citep[e.g.,][]{Jennings2012}, the results suffer from potential bias due to the presence of overlapping populations and small numbers of coeval stars.
We are interested in robustly testing the period-age relationship, and bias in the age estimates (especially from older underlying populations) could significantly affect the result.
Thus, we focus on Cepheids near high-confidence clusters, where we can expect to obtain strong age constraints.

The Andromeda Project (hereafter AP) provides an M31 cluster sample of unprecedented size and quality, derived from the PHAT survey \citep{Johnson2015}.
The PHAT data consists of HST Wide Field Camera 3 (WFC3) and Advanced Camera for Surveys (ACS) imaging in 6 filters covering $\sim 0.5$ $\mathrm{deg}^2$ of the northeast M31 disk \citep{Dalcanton2012}.
The AP clusters were identified through visual inspection of optical ACS/WFC F475W and F814W search images by citizen scientist volunteers.
We use the final AP cluster catalog of 2,753 clusters for this work.

\subsection{Cross-Matched Cluster Cepheid Sample} \label{crossmatch}

\begin{figure}
    \centering
    \includegraphics[scale=0.4]{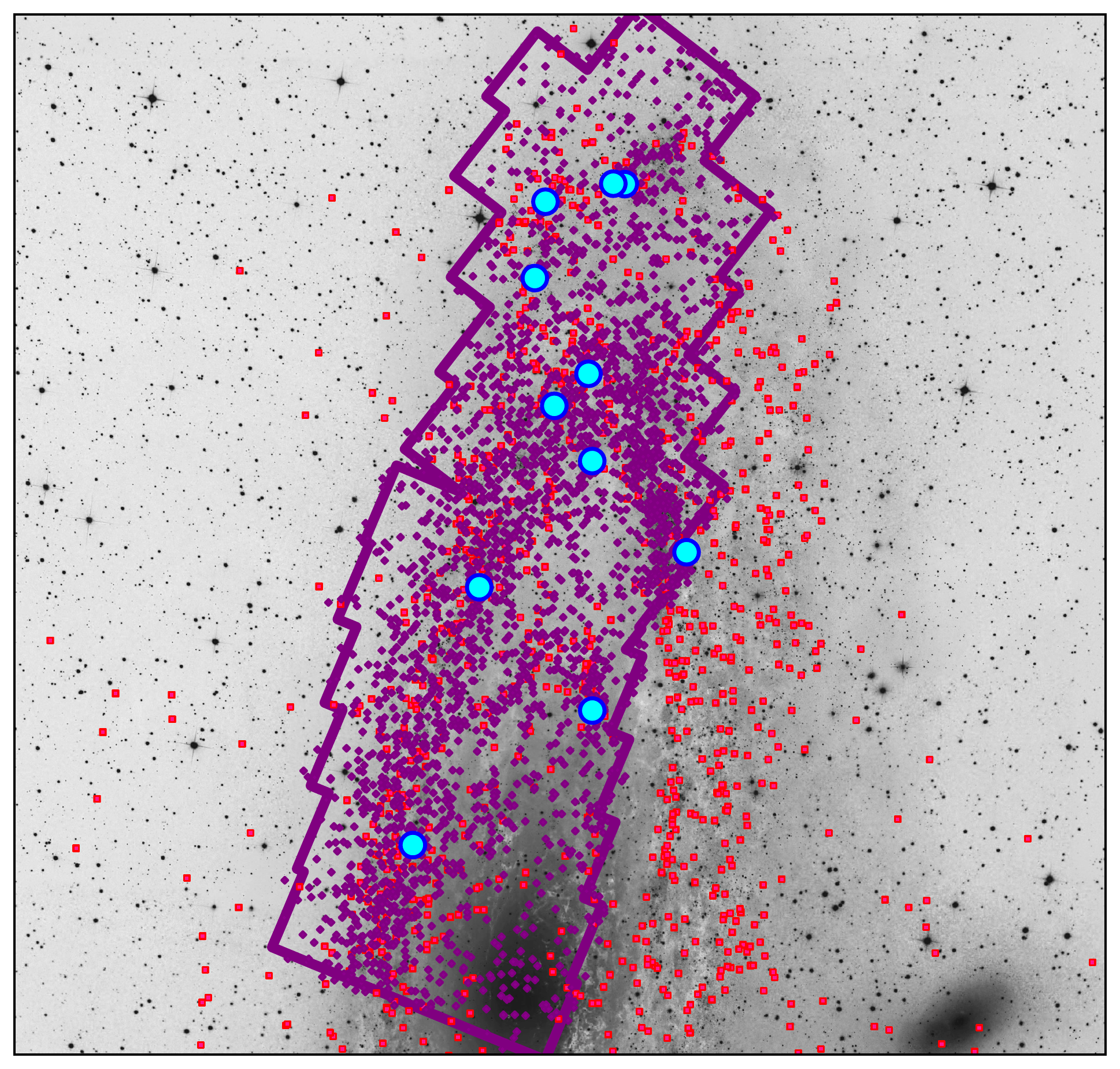}
    \caption{M31 overview; the full PHAT ACS coverage is outlined in purple, AP cluster candidates are purple diamonds, and the Kodric Cepheids are marked by red squares. The cluster Cepheids we located are denoted by large blue circles. Background image credit: Robert Gendler 
    \label{fig:m31_over}}
\end{figure}

\begin{figure}
    \centering
    \includegraphics[scale=0.4]{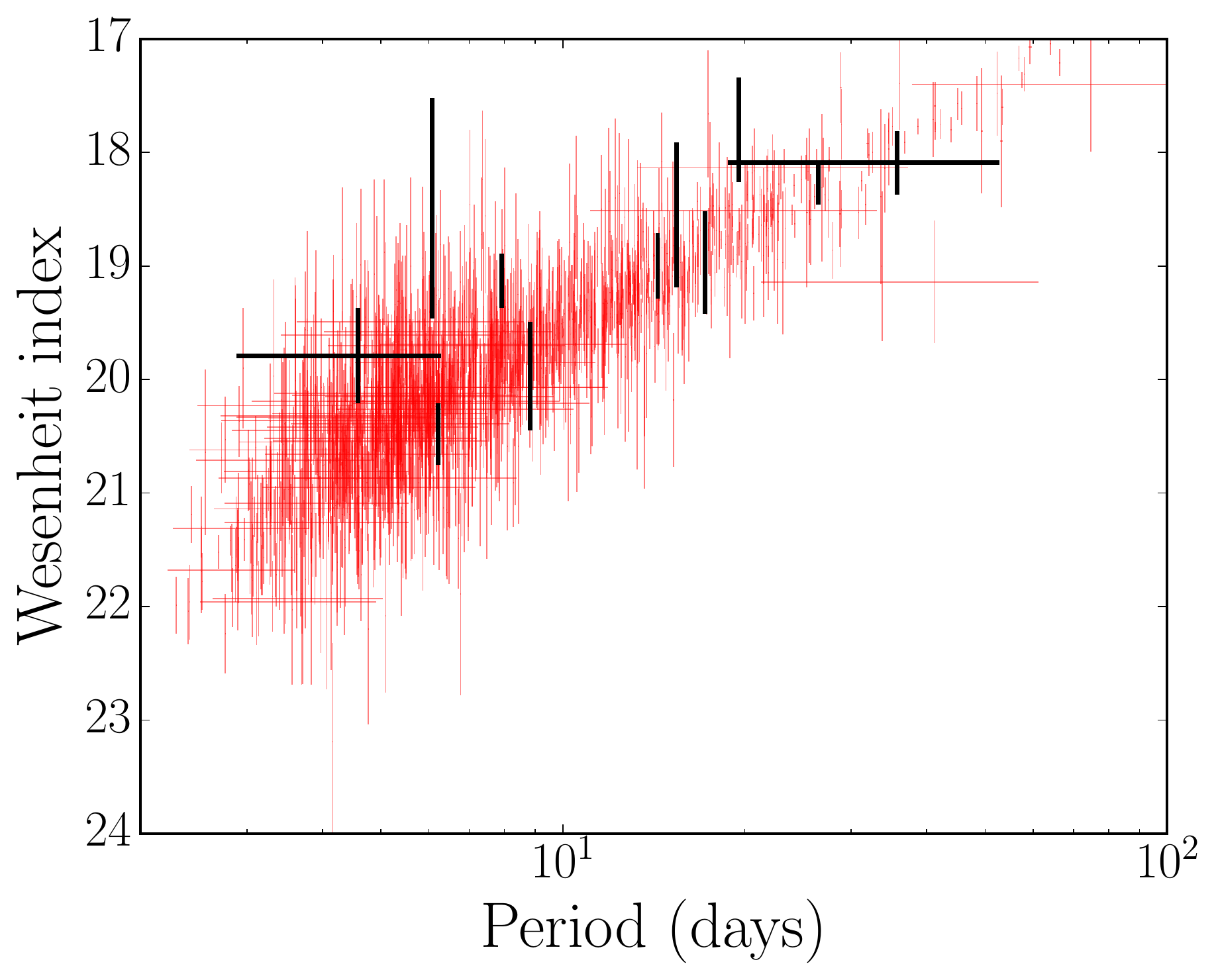}
    \caption{The Kodric fundamental-mode Cepheid sample, with the cluster Cepheid candidates in black, in period-Wesenheit space. The Wesenheit index is a color-corrected magnitude measurement intended to reduce scatter due to differential extinction \citep[in this case, derived using the \rp{} and \ip{} bands in][]{Kodric2013}. We observe that our candidates are fairly representative of the Kodric sample, though there is some indication of a bias towards brighter indices (as expected from dense cluster crowding). Note that while the uncertainties in the pulsational periods are too small to be visible on this plot for most of our sample, there are two objects with significant values.
    \label{fig:kodric_pl}}
\end{figure}

To form our sample of candidate cluster Cepheids, we first search for PAndromeda Cepheids whose reported positions place them within the extent of an AP cluster aperture.
Since the AP cluster radii and centers are somewhat uncertain, we first define the extent generously by expanding the AP radius by a factor of 1.2; however, all matches found were within the nominal cluster radii.
This search yields 9 fundamental mode (FM) Cepheids and 2 left unidentified (UN) by \citet{Kodric2013}.
Fundamental, first-overtone, and Type-II Cepheids must be studied independently for precise period-age-relations; but since all but two of our objects are firmly identified as FM pulsators and our sample size is already small, we proceed under the assumption that they are all FM Cepheids.
Table \ref{cctab} details the basic properties of this set.
The spatial and period-luminosity distribution of the sample is depicted in Figures ~\ref{fig:m31_over} and ~\ref{fig:kodric_pl}.
In addition, we note that all of the clusters selected have final weighted cluster fractions \citep[discussed in ][]{Johnson2015} $f_{clst, W}>0.6$, suggesting that all of these Cepheids fall within highly-probable clusters.

The PHAT survey implementation can be exploited to extract multi-epoch data for most points in the survey grid.
This is due to the fact that the HST tiling is based on WFC3/IR, which has a substantially smaller field of view than ACS/WFC; so simultaneous pointings of these cameras generate large overlap in the ACS images.
This means that for all but one of the Cepheids we have deep HST imagery from at least two different times, and thus a means to assess variability from the HST data alone.
This extra information can be used to distinguish the Cepheid from its neighbors at very high confidence.
The number of HST images available for the clusters under consideration ranges from 1 to 4 with a mean of 2.7.

\begin{deluxetable*}{ccccccccc}
\tabletypesize{\scriptsize}
\setlength{\tabcolsep}{0.05in}
\tablecaption{PHAT/PS1 Cluster Cepheids \label{cctab}}
\tablehead{\colhead{CC ID} & \colhead{AP ID} & \colhead{PS ID} & \colhead{ClusterFrac} & \colhead{RA (J2000)} & \colhead{DEC (J2000)} & \colhead{mode} & \colhead{Period, $r_{P1}$ (days)} & \colhead{$i_{\mathrm{P1}, HST} - i_{\mathrm{P1}, PS1}$}}
\startdata
1 & 477 & PSO-J011.4286+41.9275 & 0.9881 & 11.42865 & 41.92759 & FM & 4.582 & $1.19 \pm 0.08$ \\
2 & 1539 & PSO-J011.4279+41.8642 & 0.9358 & 11.42797 & 41.86422 & UN & 6.08 & $0.99 \pm 0.21$ \\
3 & 3928 & PSO-J011.6227+41.9637 & 0.8738 & 11.62273 & 41.96373 & FM & 6.213 & $0.09 \pm 0.06$ \\
4 & 2831 & PSO-J011.6374+42.1393 & 0.9059 & 11.63742 & 42.13933 & UN & 7.928 & $0.88 \pm 0.05$ \\
5 & 3050 & PSO-J010.9769+41.6171 & 0.8461 & 10.97692 & 41.61718 & FM & 8.829 & $0.20 \pm 0.10$ \\
6 & 5216 & PSO-J011.0696+41.8571 & 0.7049 & 11.0696 & 41.85714 & FM & 14.353 & $0.48 \pm 0.22$ \\
7 & 2113 & PSO-J011.6519+42.1286 & 0.9807 & 11.65193 & 42.12865 & FM & 15.429 & $0.11 \pm 0.13$ \\
8 & 2587 & PSO-J011.7119+42.0449 & 0.8284 & 11.71191 & 42.04491 & FM & 17.18 & $0.29 \pm 0.10$ \\
9 & 2967 & PSO-J011.0209+41.3162 & 0.6609 & 11.02092 & 41.31625 & FM & 19.566 & $0.65 \pm 0.10$ \\
10 & 1082 & PSO-J011.2797+41.6217 & 0.9087 & 11.27976 & 41.62174 & FM & 26.499 & $0.31 \pm 0.03$ \\
11 & 1540 & PSO-J011.3077+41.8500 & 0.984 & 11.30773 & 41.85003 & FM & 35.75 & $0.24 \pm 0.06$ \\
\enddata
\tablecomments{Table~\ref{cctab} lists the Cluster Cepheids (CCs) identified from the PS1/PHAT data, and provides some basic information about each. The AP ID identifies the cluster in the \citet{Johnson2015} catalog and $f_{clst, W}$ represents the final cluster quality parameter discussed there. The PS ID identifies the Cepheid from the PS catalog. The period in the \rp{} filter; Cepheid RA and Dec; the mode classification (FM for fundamental mode, UN for unidentifed) determined by \citet{Kodric2013} are also provided. Finally, the last two columns indicate the HST-determined correction to the PS1 mean magnitudes, as discussed in Section~\ref{dephase}.}
\end{deluxetable*}

\section{Analysis} \label{analysis}

Before we can study them in detail, we must confidently identify the Cepheids in the crowded PHAT fields (Section~\ref{ceph_id}).
This enables us to improve the Cepheid mean magnitudes using the superior resolution of HST (Section~\ref{dephase}); and to derive independent age estimates for the Cepheids (Section~\ref{isofits}).
The PS1 positions are particularly uncertain in clusters which HST resolves into multiple bright stars. 
In order to select the Cepheids in the HST data robustly, we examine both instability strip position \citep[using theoretical boundaries from][]{Bono2005} and PHAT field-to-field variability.

\subsection{Cepheid Identification in the HST Data} \label{ceph_id}

We first attempt to identify each Cepheid as a resolved source in the HST photometry.
Spatial agreement is checked by cluster proximity.
As the cluster fields are crowded and the PS-1 determined positions are given to an accuracy of only $0.36''$, we consider all stars within the AP cluster radius as candidates.
We consider two additional pieces of evidence when cross matching the Cepheids; location on the CMD and variability.

\emph{Proximity to the instability strip:}
We adopt the ``non-canonical", solar-metallicity instability strip boundaries derived by \citet{Bono2005} for fundamental pulsators satisfying $0.17 < \log_{10}(\mathrm{period}/\mathrm{days}) < 2.02$.
In order to transform the boundaries from theoretical $T_{\mathrm{eff}}$-Luminosity space to HST filter space, we use the MATCH \citep{Dolphin2002} \texttt{fake} utility to sample the Padova isochrones (discussed in Section~\ref{isofits}) at various ages and thereby populate the instability strip with model stars.
The model stars which fall between the blue and red edges and within the valid period range \citep[determined using the pulsational relations derived in][]{Bono2000} are then plotted in HST F475W-F814W, F814W color-magnitude space.
The quadrilateral which most-tightly encloses these points in CMD space defines our observational instability strip.
Indeed we have used the transformations from \citet{Sirianni2005}, as described in \citet{Girardi2008} (as these transformations are used by MATCH to yield filter magnitudes).

\begin{figure*}
    \centering
    \includegraphics[scale=0.6]{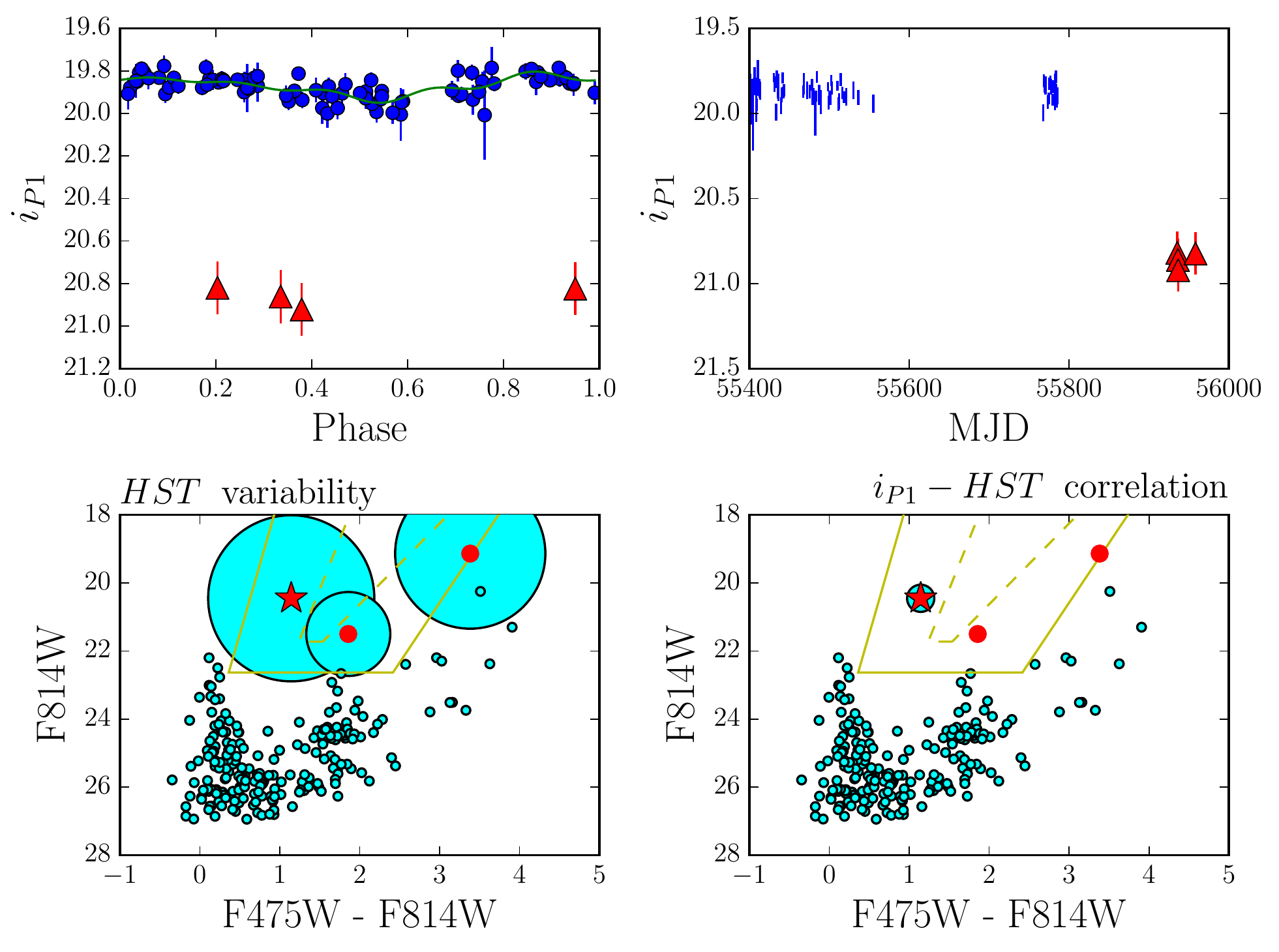}
    \caption{Details of the process of Cepheid identification in the HST data for CC2. Top row: (left) the Kodric lightcurve and model fit (blue circles and green line, respectively) with filter-transformed and dephased HST data (red triangles; see Section~\ref{dephase}); (right) same, plotted against time (MJD). Note the large offset between the PS1 and HST magnitudes --- this difference is discussed in terms of blending in Section~\ref{dephase}. Bottom row: (left) the HST cluster color-magnitude diagram with each star marked by a cyan circle. The size of the cyan circles scales with the value of the HST F475W-F814W correlation metric discussed in Section~\ref{ceph_id} (stars with very small values are given a fixed point size) - candidate instability strip members are additionally marked by red dots, and the putative Cepheid by a red star - the instability strip (original, yellow dashed line; and broadened, yellow solid line) are outlined; (right) same as bottom-left, with the HST-\ip{} correlation metric (Section~\ref{ceph_id}). Here, we see that one object within the instability strip stands out in both the HST variability metric and the HST-\ip{} correlation metric (the largest cyan circle in both of the bottom plots). The remaining images in this series are presented in the Appendix.}
    \label{fig:cmdvar}
\end{figure*}

Membership in the final filter-transformed instability strip is uncertain due to errors in the reddening adopted for each cluster.
We estimate \av{} by fitting the CMD of the full cluster (as described in Section~\ref{isofits}).
The uncertainty in the reddening estimates from these fits is on the order of 0.1 mag.
To ensure that we catch all reasonable candidates, we expand the edges of the strip to cover an \av{} range of $\pm 1.5$ mag around the best fit, as illustrated in Figure~\ref{fig:cmdvar}.
This is a very large overestimate intended to prevent accumulated error from the HST photometry, the filter corrections, and pulsational models from excluding too many objects from consideration.

\emph{Variability:}
For each cluster, we identify all ACS observations from the PHAT survey whose borders contain the cluster.
We select a reference field in which the cluster is fully contained and that is not too close to the chip gap (where distortion potential is highest).
We start with the \texttt{st} photometry catalogs \citep[described by][]{Williams2014}, and apply an additional cut in S/N (greater than 4 in both filters).
These catalogs are produced by the PHAT reduction pipeline, and do not include quality cuts on PSF shape or crowding; we wish to maximize catalog completeness at the expense of photometry quality.
In order to match the stars in different overlapping observations, we use pattern-matching code from \citet{Groth1986} to determine RA/Dec offsets from the brightest stars in each field.
For each catalogued source in the reference field, we search a circle of radius .072 arcseconds in the other images for matches.
This radius is on the order of the FWHM for ACS, and our results indicate that it matches all sources of interest.
If more than one source is found in any of these circles, we consider it a non-detection in the corresponding field.
The sources for which we obtain at least 2 exposures are retained for analysis.
We compare the resulting catalog to the full \texttt{st} reference field catalog in the instability strip region to ensure that no candidates are missed due to alignment issues.

Before we can compare the PHAT and PS1 data, we must extend the PS1 lightcurves to the HST observation epochs.
For each Cepheid in their final sample, \citet{Kodric2013} provide the raw unfolded lightcurves and the coefficients of a fifth-order Fourier series fit to the folded lightcurves.
Unfortunately, they do not appear to provide a zero-epoch for these determinations.
We use a simple scalar minimization routine to determine the zero epoch which leads to the best agreement between the Fourier series and the data.
The result is checked by-eye, and the uncertainty in this epoch determination is computed by bootstrap resampling.
This procedure allows us to compute the Cepheid magnitude expected by PS1 at an arbitrary epoch.
Uncertainties in the PS1 expected magnitudes are propagated by Monte Carlo methods, assuming that the parameters and associated errors provided by Kodric represent Gaussian distributions.

We compute several variability indices inspired by \citet{Welch1993} for each matched source.
These quantify the degree to which the magnitudes in two filters are correlated from field to field.
They are calculated for our pair of HST ACS magnitudes F475W and F814W for each source, as well as for the source F814W and the model expected \ip{} Cepheid magnitude.
We denote filter 1, filter 2 magnitude in the \(k^{th}\) field by \(F1_k, F2_k\) (respectively), the corresponding pipeline-determined Poisson uncertainty in each by \(\sigma_{f1, k}, \sigma_{f2, k}\), and the variance-weighted mean for each filter by $\overline{F1}$.
We then define the variability index \(I\) to be
\[ I = \frac{1}{\sqrt{n(n-1)}} \sum_{i=1}^n  \left(\frac{F1_k - \overline{F1}}{\sigma_{f1,k}}\right) \left(\frac{F2_k - \overline{F2}}{\sigma_{f2,k}}\right)\]
where \(n\) is the number of fields in which the source was matched.
We denote the index computed with the 2 ACS filters as \Iacs{}; and with the F814W and expected \ip{} as \Iacsip{}.
Note that this index is the covariance of the measurements normalized by the photometric uncertainty.
For a truly non-variable source with good photometry and good background subtraction, the variation in measured brightness in two filters should be independent.
Thus the expectation value of \(I\) for non-variable stars with ideal photometry is zero.
A higher value of \(I\) indicates a higher degree of correlation between brightness changes in the two filters.
A high value of the index may indicate a truly variable source; but it can also reflect issues with photometry, background subtraction, or source matching.
Here, our situation is simplified - we expect that there is a variable within the field being examined, and the indices are used to help us identify which candidate resolved source it is most likely to be.

We select as candidates the sources whose weighted-average ACS magnitudes place them inside of the generous instability strip for the cluster (yellow-outlined region in Figure~\ref{fig:cmdvar}).
We identify the putative Cepheid as the candidate source with the maximum variability index, considering both filters equally.
Since the photometric errors of the ACS observations are substantially smaller than those assigned to the PS1 magnitudes by our Monte Carlo uncertainty estimation, the ACS variability index usually dominates (especially when the brightness changes are slight).
This selection process is illustrated for each cluster in Figure~\ref{fig:cmdvar}.
For all clusters except AP5216, we find a clear variability standout in the expected region of the CMD.
Since AP5216 was observed in only one PHAT field, no variability information is available for the HST sources; fortunately, the generous instability strip encloses exactly one cluster source, which we identify as CC6.

\subsection{Cepheid Dephasing} \label{dephase}

We now proceed to compare the HST and PS1 photometry of the Cepheids.
The cluster environments are crowded, so we expect the PS1 measurements to be systematically brighter due to source confusion.
With our Cepheids identified in the HST data, we can test the actual size of this effect.
In this section, we use the absolute flux estimate from the HST photometry to re-estimate the Cepheid mean magnitudes.
This requires that we compare the HST photometry to the PS1 models in a common filter system.

Since we are only concerned with Cepheids, whose SEDs are fairly homogeneous and not unusual, a standard color-based transformation is appropriate here.
Both the F814W ACS filter and the \ip{} filter are close to the Landolt (Johnsons/Cousins) $I$.
\citet{Tonry2012} has conducted an analysis of the Pan-STARRS1 photometric system, which includes transformations to the Johnsons/Cousins system derived from a library of 783 photometric-standard SEDs.
Similarly, \citet{Sirianni2005} presents transformations from ACS magnitudes to Johnsons/Cousins based on synthetic magnitudes and checked with empirical data.
We opt to use the quadratic relations in both cases.
After de-reddening the HST data using our cluster \av{}, we transform each pair of F475W/F814W magnitudes to $V/I$ by iteratively improving the $V-I$ color term.
The PS1 transformations require we use $B-V$ color.
Since we are missing a coincident $B$ observation, we opt to estimate the Cepheid's average $B-V$ color using an empirical period-color relation from \citet{Tammann2003} for galactic Cepheids.
To account for the variation of $B-V$ over the period of the Cepheid, we add 0.2 mags in-quadrature to the color uncertainty.
The reddening is re-applied to the final magnitudes using the prescriptions from \citet{Tonry2012}.
So for each field in which the Cepheid candidate appeared, we obtain an HST-derived estimate of the Cepheid brightness in the \ip{} filter at that field epoch, which we can compare to the PS1 model magnitude.

Now we use the HST data to cross-calibrate the PS1 lightcurves.
The PS1 magnitude variation provides a good estimate of the true change in flux received from the Cepheid, but the absolute level of the flux is better determined by HST.
At each HST epoch, we compute the difference in zeropoint-scaled flux between the PS1-determined mean magnitude and the PS1 model magnitude at that time:
\[ \Delta f = 10^{-0.4 \bar{i}_{P1}} - 10^{-0.4 i_{ \text{P1, model} }}   \]
then add this to the HST-determined magnitude in the \ip{} filter:
\[ \bar{i}_{P1, \text{HST}} = -2.5 \log_{10}(10^{-2 i_{P1, HST}/5} + \Delta f). \]
We repeat this for each field, and propagate the uncertainties via Monte Carlo methods (assuming Gaussian distributions for the parameters).
Thus, for each Cepheid, we obtain at least 2 estimates of a calibrated mean magnitude in \ip{}.

\begin{figure}
    \centering
    \includegraphics[scale=0.4]{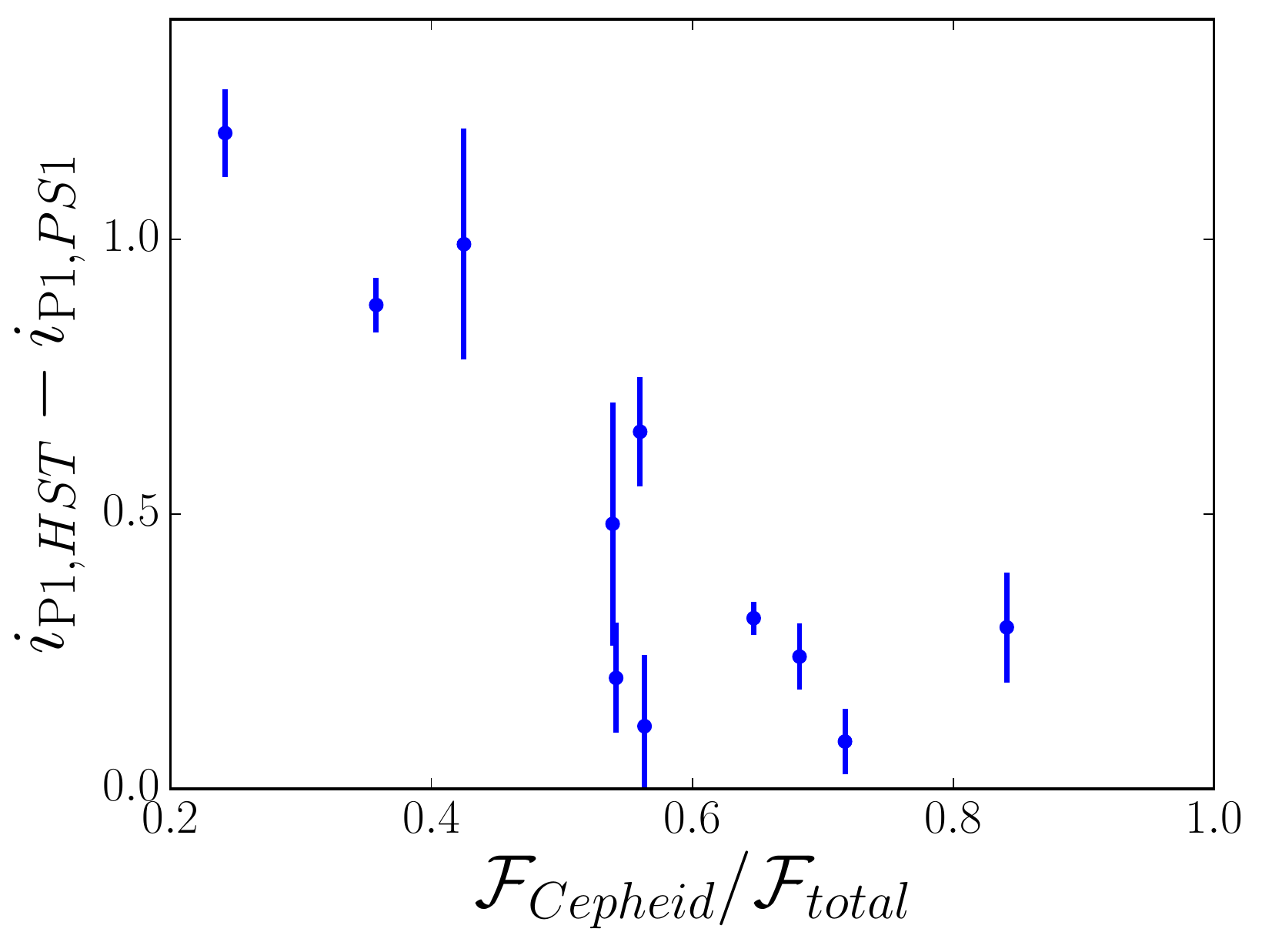}
    \caption{The difference in \ip{} magnitude between the HST dephased means and those determined by PS1, plotted against a blending metric. The metric $\mathcal{F}_{Cepheid}/\mathcal{F}_{total}$ is the ratio of the putative Cepheid flux to that of all HST-resolved and cataloged sources within 0.86". The results from each PHAT field are averaged. There is a clear correlation between field crowding and brighter PS1 magnitudes.}
    \label{fig:dephased_fluxcomp}
\end{figure}

The results of this dephasing step are shown in Figure~\ref{fig:dephased_fluxcomp}.
We plot the difference between the HST-calibrated mean magnitude and the PS1 mean magnitude (averaging the results for all PHAT fields the star was located in), against a blending measure.
The measure is computed by summing the F814W flux of all \texttt{.dst} catalog sources within a radius of 0.86'' (3.3 parsecs in the disk) of the putative Cepheid.
In describing the PAndromeda data pipeline, \citet{Lee2012} reports for one representative skycell a median FWHM in the 30 best images of $0.861''$.
Thus, the metric represents a rough estimate of the degree to which blending affects the PS1 data.
The HST-calibrated means are systematically dimmer than the PS1 means, and there is a trend towards better agreement with less excess (HST-cataloged) flux near the Cepheid.

As expected in crowded cluster environments, the PS1 magnitudes appear to be systematically brightened by blending.
This effect has been studied in-depth previously, e.g. in relation to the DIRECT project by \citet{Mochejska2000} and by \citet{Vilardell2007}.
In finding a PL relation for their sample, \citet{Kodric2013} performs iterative $3\sigma$ clipping to eliminate blends and other outliers.
We repeat their procedure, with both the full set of Kodric FM and UN Cepheids; and with the FM Cepheids alone.
In both cases, only one Cepheid from our cluster Cepheids sample (CC1) is cut (after 5 iterations in the former, and on the first iteration in the latter case).
This leaves 2 (4, including the UN) Cepheids in the PL-determining set whose PS1 measured magnitudes are offset at the 0.5-1.0 mag level.

\begin{figure}
    \centering
    \includegraphics[scale=0.4]{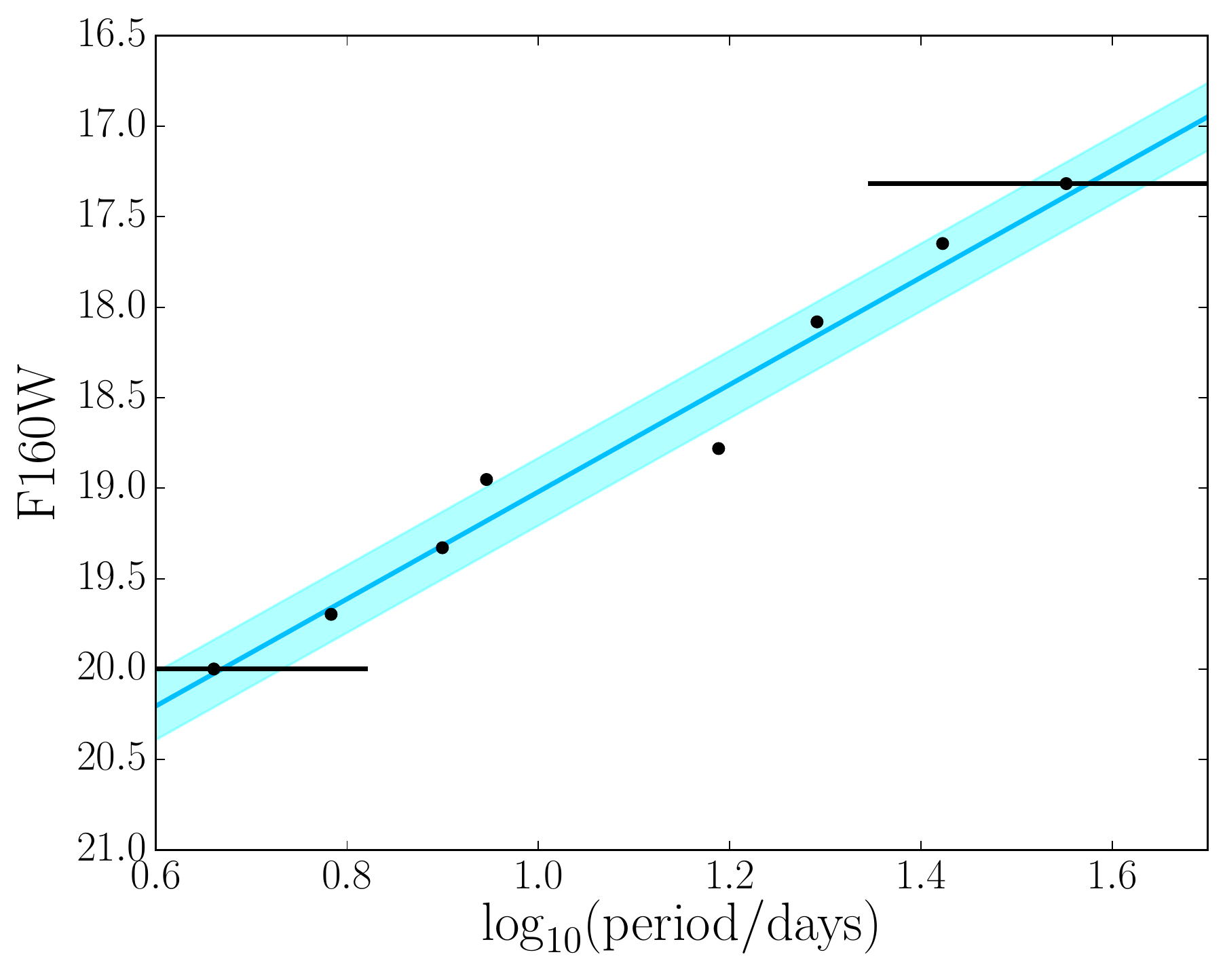}
    \caption{The F160W PL relation for our cluster Cepheids, compared with the empirical linear PL relation found by \citet{WagnerKaiser2015} in M31. The two agree within the uncertainty, which is represented by the filled section; the magnitude errors in our data are photometry-derived only (no consideration was given to Cepheid phase). Three Cepheids were excluded due to NIR coverage issues. Note that the uncertainties in the Cepheid magnitudes and periods are smaller than the point size on this plot except for two of the period measurements.}
    \label{fig:nir_pl}
\end{figure}

We also find the Cepheid magnitudes in the near-infrared filter F160W.
The effects of random-phase observations are diminished substantially in this filter, as exploited in M31 by \citet{Riess2012}, \citet{Kodric2015}, and \citet{WagnerKaiser2015}.
The F160W magnitudes of 8 of our cluster Cepheids (3 were not imaged in NIR filters) are plotted in Figure~\ref{fig:nir_pl}.
The empirical (linear) F160W PL relation derived by \citeauthor{WagnerKaiser2015}\ is also plotted.
Our Cepheids show good agreement with this relationship within the measured uncertainty.

\subsection{Cluster Age Determination} \label{isofits}

We estimate the age of the host stellar clusters by fitting populated isochrones to their CMDs.
In particular, we use Padova isochrones and assume a Kroupa IMF.
These isochrones combine the tracks from \citet{Girardi2000} through the first thermal pulse on the AGB with tracks that treat the TP-AGB in better physical detail, as described in \citet{Marigo2008}.
Corrections from \citet{Girardi2010} concerning the AGB are also incorporated.
In this paper, we are concerned with young Cepheid-hosting clusters only, so the TP-AGB/AGB corrections are not critical.
The Padova isochrones are computed over a large age range (from 6.60 to 10.10 dex in $\log(t/\text{years})$) which should comfortably capture all FM Cepheids.
The Padova models use a moderate amount of convective overshoot.

We fit the cluster CMDs using MATCH \citep{Dolphin2002}.
We run MATCH in a mode that fits a single stellar population (SSP) to the cluster CMD.
MATCH generates model CMDs which include the effects of cluster-specific artificial star tests, and thus has some defense against spurious photometry effects in these crowded fields.
The methodology, fitting parameters, and data products (photometry, artificial star tests, etc.) are the same as those used in the (forthcoming) CMD analysis of the full PHAT cluster sample (L. Beerman et al.\, in preparation).
MATCH accounts for potential field star contamination in the cluster CMD by incorporating a smooth binning of likely field stars situated in an annulus 10 times the size of the cluster aperture in the fit.
We fix the cluster distance modulus at $24.47$, and allow the metallicity of the cluster to vary from $\log_{10} (Z_{\mathrm{min}}) = -0.20$ to $\log_{10} (Z_{\mathrm{max}}) = 0.10$ in steps of $0.1$ dex.
The runs are performed with a resolution in $\log_{10}(t/\mathrm{years})$ of $0.10$ over the range $(6.6, 10.1)$; and a resolution in \av{} of $0.05$ over the range $(0.00, 3.00)$.
These ranges encompass all physically-reasonable solutions, assuming the Cepheids are cluster members.
By fitting all models in this grid, and marginalizing over \av{} and metallicity, we obtain a posterior probability density function (PDF) for the cluster age assuming the model we've adopted.

We perform the fits with and without the Cepheid included in the data.
This serves several purposes.
First, isochrone fits are most robust (and degeneracy is minimized) when they are determined by main sequence stars rather than a handful of bright evolved stars.
This is especially worth considering in these cases, where the Cepheid is often the only star in the corresponding section of the CMD.
Secondly, this enables us to perform a meaningful test of the period-age relation.
If the CMD fits are controlled by the Cepheid photometry, then the resulting MATCH age estimate is effectively a measurement of the Padova isochrone age at a given blue loop magnitude.
Finally, since we are not certain that the Cepheid belongs to a dominant SSP as we have assumed, we can look for large changes in the PDF when the Cepheid is removed, which might indicate that this assumption fails.
The reported ages used in Section~\ref{discussion} are those derived from the CMDs without the Cepheid.

In most cases, the PDFs with and without the Cepheid are very similar.
Six clusters display differences in the best-fit age at our $0.10$ dex resolution, and three of these show differences in the median age estimates.
We can produce a rough approximation of the standard error of these estimates from half the 84-16 percentile difference.
Only one cluster (CC2) shows a difference in both best-fit and median age which exceeds the joint errors. 
The actual shift in both age estimates is only 0.20 dex.
Several other PDFs show hints of a bimodal shift when the Cepheid is removed.
Qualitatively, the lack of large shifts in the fitting results indicate that the MATCH fits are indeed capturing an underlying SSP which is roughly consistent with the Cepheids.

The likelihood of the best-fit SSP model relative to the null model (no SSP, i.e. based only on the empirical field star CMD) is another test of fit quality.
We re-run MATCH on each cluster, but constrain it to use only the smoothed background model derived from the cluster annulus.
We are concerned with the quantity $\Delta f = f_{\mathrm{best}}-f_{\mathrm{background}}$ where $f$ is the fit value MATCH returns.
This $f$ is a Poisson-statistics analogue to $\chi^2$, and thus roughly corresponds to $-2 \ln(P)$; so the quantity $e^{-\Delta f /2} \sim P_{\mathrm{best}}/P_{\mathrm{background}}$ informs us how much more likely the best fit found is relative to the null background fit.
This likelihood ratio ranges from $10^3$ to $10^{28}$ for our cluster fits without the Cepheid.
The outliers at the low end of this quantity have correspondingly broad normalized PDFs, by necessity.
We conclude that there are no catastrophic failures in the fitting results, and proceed with the full cluster age PDFs (derived without the Cepheids) in hand.

\section{Results} \label{discussion}

We now use our period measurements and cluster age estimates to test the period-age relation for Cepheids in M31.
This test requires a robust analysis of our MATCH results and a set of model period-age relations to compare with.

\subsection{Fitting Our Period-Age Data} \label{par_fit}

We adopt a probabilistic approach to fitting our data.
This enables us to properly incorporate the full (non-Gaussian) probability distributions we computed using MATCH for the cluster ages, and to account for errors in the period determination from \cite{Kodric2013} (substantial in two cases, as shown in Figure~\ref{fig:kodric_pl}).

We fit a linear model to the period and age data: $\log_{10}(\mathrm{age/years}) = m\log_{10}(\mathrm{period/days}) + b$, with  Gaussian scatter $\sigma$ in magnitude about the line.
Rather than fit with $m$ directly, we use $\theta = \arctan(m)$ so that a flat prior doesn't favor steeper slopes.
Given our data (represented by $D$) and our SSP model for the clusters ($\Pi$), we would like to determine the probability function for a linear slope and intercept: $\prob(\set{\theta, b, \sigma} \given D,\Pi)$.
Baye's rule informs us that
\[ \prob(\set{\theta, b, \sigma} \given D,\Pi) \propto \prob(D \given \set{\theta,b,\sigma}, \Pi) \; \prob(\set{\theta,b,\sigma}) \]
where $\prob(\set{\theta,b,\sigma})$ represents our prior expectation for the fit parameters.
We adopt a very generous prior (flat over $(-\pi/2<\theta<0)$ and all possible intercepts).
To evaluate the likelihood function $\prob(D \given \set{\theta,b,\sigma}, \Pi)$, we integrate the product of 1) the MATCH-provided PDFs, 2) the Gaussian period PDFs, and 3) the modeled Gaussian variance about the line $(\theta,b,\sigma)$ over a grid of points for each Cepheid.
The product of these likelihoods for all the Cepheids yields the full likelihood function.
We apply Markov chain Monte Carlo (MCMC) \citep[using the package \texttt{emcee}:][]{Foreman2013} to sample the resultant posterior distribution.

The MCMC fit is displayed in Figure~\ref{fig:pa_fit}.
Using the 16th/50th/84th percentiles of the MCMC chain, we obtain estimates for the slope and intercept of $m = -0.69^{+ 0.25}_{- 0.20}$, $b = 8.38^{+ 0.37}_{- 0.36}$; and for the scatter, $\sigma = 0.25^{+0.33}_{-0.20}$.
The constraints are broad, as can be seen in the wide range of the sampled lines in Figure~\ref{fig:pa_fit}.
This is a product of the broad age PDFs, which in turn indicate the difficulty of assigning an age to low-mass clusters at the distance of M31.

\subsection{Comparison and Discussion} \label{par_comparison}

We compare our results to several period-age relations in the literature.
It is important to note that the choice of evolutionary model used (in fitting clusters or modeling Cepheids) will in general affect the period-age relation, as it affects the age-luminosity relation for stars near the instability strip.
From \citet{Bono2005}, we compare with the classical, fundamental-mode, relation, for $Z=0.02$ (solar metallity).
This relation was derived using stellar models from \citet{Pietrinferni2004} which do not include convective core-overshooting.

It is more difficult to find period-age relations derived using models which incorporate overshooting, such as the Padova models we used for CMD fitting.
Thus, we derive our own semi-empirical period-age relation from the Padova models.
Specifically, we use the MATCH utility \texttt{fake} to sample the Padova isochrones in the range $6.6<\log(\mathrm{age/years})<8.5$, at a resolution of 0.05 dex.
We adopt a sufficiently high SFR to populate all instability strip crossings.
All other parameters (except initial distance modulus) are kept identical to their state in the CMD fitting (Section~\ref{isofits}).
We select the models lying within the $T_{eff}$, luminosity boundaries derived by \citet{Bono2005} (solar metallicity, non-canonical).
We then utilize the theoretical pulsation relations for fundamental Cepheids derived by \citet{Bono2000} to relate the model temperatures, luminosities, and masses to pulsational periods.
Then a simple linear fit is performed to the resulting points in period-age space, providing the desired theoretical prediction.

\begin{figure}
    \centering
    \includegraphics[scale=0.4]{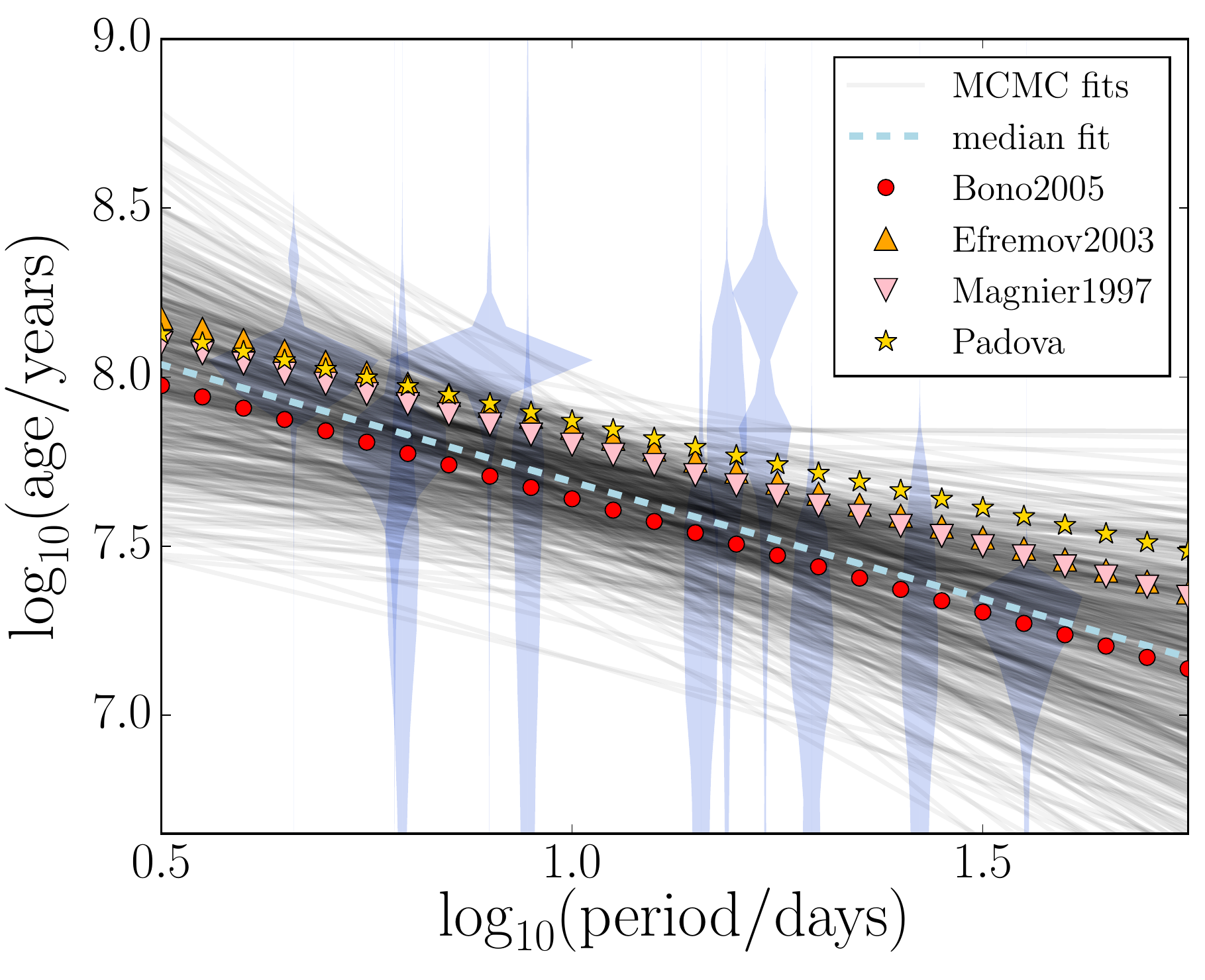}
    \caption{The period-age relation fit for our cluster Cepheids. The blue violin plots represent the age PDFs produced with MATCH; their width at each age value represents the likelihood of that age from the MATCH SSP model. The period error is not marked on this plot (see Figure~\ref{fig:kodric_pl}). The gray lines are random samples from the MCMC chain of fits. The line marked by yellow triangles is the Padova period-age relation derived in Section~\ref{par_comparison}; the red circle line represents the solar-metallicity period-age relation derived by \citet{Bono2005}; the orange-upper and pink-lower triangles represent the period-age relations found by \citet{Efremov2003} and \citet{Magnier1997}, respectively.}
    \label{fig:pa_fit}
\end{figure}

For further comparison, we locate two additional relations from the literature.
\citet{Magnier1997} calculate a period-age relation in a semi-empirical way: they use tracks from \citet{Schaller1992} which allow for overshooting, and an empirical period-luminosity relation found for M31 Cepheids.
And \citet{Efremov2003} fit a period-age relation empirically to a sample of Cepheids from the LMC bar, utilizing integrated color results based upon models from \citet{Bertelli1994} which do incorporate moderate convective core overshooting; we quote their most probable relation.
We plot these period-age relations in Figure~\ref{fig:pa_fit} alongside our MCMC fits, and compare the fit parameters in Table~\ref{tab:pa_fit}.
The median estimates agree reasonably well with the ensemble of other values given the large uncertainties in our fit.

It is interesting to note that the largest disagreement in slope value is between our empirical fit and the Padova model predictions we derived, despite the fact that both rely on the same set of stellar models.
To check whether this discrepancy is due to our choice of pulsational models, we also applied the approach of \citet{Magnier1997} and used empirical period-luminosity relations to relate the theoretical model luminosities to pulsational periods and re-derive a Padova period-age prediction.
The MATCH \texttt{fake} utility provides magnitudes in common filter systems for the model stars \citep{Girardi2008}.
We re-derive the period-age relation using several different period-luminosity relations: two from \citet{WagnerKaiser2015} for the NIR HST filter F160W and for the I-band; and two from \citet{Fouque2007} derived using galactic Cepheids in the V and I bands.
The slopes derived using \citet{WagnerKaiser2015} are close to the shallow theoretical Padova results described above: $m \simeq -0.55$ for both F160W and I.
In contrast, the slopes found using \citet{Fouque2007} are closer to our steeper empirical fit: $m \simeq -0.78, -0.69$ for V, I respectively.
We expect that the results from \citet{WagnerKaiser2015} will be more robust, since their sample size is larger and Cepheid variability is reduced at long wavelengths.
This reinforces the suggestion from the theoretical Padova fit that we should expect a shallower slope.
Confirming this experimentally would require further testing on clusters with better constraints than our M31 sample can provide.

We conclude that our Cepheid age measurements are in broad agreement with existing theoretical and empirical period-age relations.
There are several important caveats to note.
Comparing ages derived using stellar models with differing assumptions about overshooting is not strictly valid, as overshooting parameters have an effect on the derived age at fixed luminosity.
In addition, we have compared period-age relations derived at different metallicities (solar and LMC/SMC).
Our constraints on the period-age relation are too loose to distinguish these second-order effects; but they must be recognized when applying these tools to derive age estimates.

\begin{deluxetable}{cccc}
\tabletypesize{\scriptsize}
\setlength{\tabcolsep}{0.05in}
\tablecaption{Period-Age Relations $\log_{10}(\mathrm{age/years}) = m\log_{10}(\mathrm{Period/days}) + b$ \label{tab:pa_fit} }
\tablehead{\colhead{Source} & \colhead{$m$} & \colhead{$b$} & \colhead{Scatter}\tablenotemark{a}}
\startdata
M31 CC fit & $-0.69^{+ 0.25}_{- 0.20}$ & $8.38^{+ 0.37}_{- 0.36}$ & $0.25^{+ 0.33}_{- 0.20}$ \\
Padova fit & $-0.53$ & $8.40$ & $0.07$ \\
\citet{Bono2005} & $-0.67$ & $8.31$ & $0.08$ \\
\citet{Magnier1997} & $-0.6$ & $8.4$ & -- \\
\citet{Efremov2003} & $-0.65$ & $8.50$ & --
\enddata

\tablenotetext{a}{Predicted scatter in magnitudes about the best-fit relation.}
\end{deluxetable}

\section{Conclusion}
\label{conclusion}
In this paper, we presented and analyzed a set of 11 Cepheids in M31 which appear to be members of partially-resolved clusters.
The sample was found by cross-correlating the M31 Cepheid sample from \citet{Kodric2013} with the PHAT cluster sample presented by \citet{Johnson2015}.
We located the Cepheids in the HST photometry by their variability and CMD locations.
This revealed that the PS1 mean magnitudes were systematically brightened by crowding.
We then performed fits to the selected cluster CMDs without the Cepheids.
Our derived period-age relation for these objects is poorly-constrained, but agrees broadly with the body of existing relations in the literature.

The sample of cluster Cepheids we present are particularly well-characterized for objects at the distance of M31.
Combining the HST and PS1 photometry allows the former to be properly dephased and the latter to be crowding-corrected.
Our magnitude calibration results indicate that the PS1 mean magnitudes are biased bright at the 1 magnitude level in the highest stellar density regions.
\citet{WagnerKaiser2015} uses PHAT photometry to further explore the effect of such contamination on the period-luminosity relation for M31 Cepheids.

Cluster Cepheids are essential calibrators for period-age relations and stellar evolution generally.
This work constitutes the first direct test of the period-age relation by CMD-fitting beyond the Galaxy and Magellanic clouds.
Our results support application of such period-age relations to M31 and other solar-metallicity galaxies.
Further investigation of the full PS1 Cepheid sample in this context, once extended to more of the M31 halo and to larger periods \citep{Kodric2013}, could provide additional insight into the recent star formation history of M31.

\acknowledgements
{We would like to thank the anonymous referee for a thorough and helpful report. This work was supported by the Space Telescope Science Institute through GO-12055. P.S. acknowledges financial support from the Washington NASA Space Grant Consortium, NASA Grant \#NNX10AK64H. This research made use of NASA's Astrophysics Data System; APLpy, an open-source plotting package for Python hosted at http://aplpy.github.com; Astropy, a community-developed core Python package for Astronomy \citep{2013A&A...558A..33A}; the IPython package \citep{PER-GRA:2007}; matplotlib, a Python library for publication quality graphics \citep{Hunter:2007}.
}

\bibliography{./phatceph.bib}

\begin{thebibliography}{}
\expandafter\ifx\csname natexlab\endcsname\relax\def\natexlab#1{#1}\fi

\bibitem[{Anderson {et~al.}(2013)Anderson, Eyer, \& Mowlavi}]{Anderson2013}
Anderson, R.~I., Eyer, L., \& Mowlavi, N. 2013, MNRAS, 434, 2238

\bibitem[{{Astropy Collaboration} {et~al.}(2013){Astropy Collaboration},
  {Robitaille}, {Tollerud}, {Greenfield}, {Droettboom}, {Bray}, {Aldcroft},
  {Davis}, {Ginsburg}, {Price-Whelan}, {Kerzendorf}, {Conley}, {Crighton},
  {Barbary}, {Muna}, {Ferguson}, {Grollier}, {Parikh}, {Nair}, {Unther},
  {Deil}, {Woillez}, {Conseil}, {Kramer}, {Turner}, {Singer}, {Fox}, {Weaver},
  {Zabalza}, {Edwards}, {Azalee Bostroem}, {Burke}, {Casey}, {Crawford},
  {Dencheva}, {Ely}, {Jenness}, {Labrie}, {Lim}, {Pierfederici}, {Pontzen},
  {Ptak}, {Refsdal}, {Servillat}, \& {Streicher}}]{2013A&A...558A..33A}
{Astropy Collaboration}, {Robitaille}, T.~P., {Tollerud}, E.~J., {et~al.} 2013,
  \aap, 558, A33

\bibitem[{Bertelli {et~al.}(1994)Bertelli, Bressan, Chiosi, Fagotto, \&
  Nasi}]{Bertelli1994}
Bertelli, G., Bressan, A., Chiosi, C., Fagotto, F., \& Nasi, E. 1994, A\&AS,
  106, 275

\bibitem[{{Bono} {et~al.}(2000){Bono}, {Castellani}, \& {Marconi}}]{Bono2000}
{Bono}, G., {Castellani}, V., \& {Marconi}, M. 2000, \apj, 529, 293

\bibitem[{Bono {et~al.}(2005)Bono, Marconi, Cassisi, Caputo, Gieren, \&
  Pietrzynski}]{Bono2005}
Bono, G., Marconi, M., Cassisi, S., {et~al.} 2005, ApJ, 621, 966

\bibitem[{Dalcanton {et~al.}(2012)Dalcanton, Williams, Lang, Lauer, Kalirai,
  Seth, Dolphin, Rosenfield, Weisz, Bell, Bianchi, Boyer, Caldwell, Dong,
  Dorman, Gilbert, Girardi, Gogarten, Gordon, Guhathakurta, Hodge, Holtzman,
  Johnson, Larsen, Lewis, Melbourne, Olsen, Rix, Rosema, Saha, Sarajedini,
  Skillman, \& Stanek}]{Dalcanton2012}
Dalcanton, J.~J., Williams, B.~F., Lang, D., {et~al.} 2012, ApJS, 200, 18

\bibitem[{Dolphin(2002)}]{Dolphin2002}
Dolphin, A.~E. 2002, MNRAS, 332, 91

\bibitem[{Efremov(2003)}]{Efremov2003}
Efremov, Y.~N. 2003, Astronomy Reports, 47, 1000

\bibitem[{{Foreman-Mackey} {et~al.}(2013){Foreman-Mackey}, {Hogg}, {Lang}, \&
  {Goodman}}]{Foreman2013}
{Foreman-Mackey}, D., {Hogg}, D.~W., {Lang}, D., \& {Goodman}, J. 2013, \pasp,
  125, 306

\bibitem[{{Fouqu{\'e}} {et~al.}(2007){Fouqu{\'e}}, {Arriagada}, {Storm},
  {Barnes}, {Nardetto}, {M{\'e}rand}, {Kervella}, {Gieren}, {Bersier},
  {Benedict}, \& {McArthur}}]{Fouque2007}
{Fouqu{\'e}}, P., {Arriagada}, P., {Storm}, J., {et~al.} 2007, \aap, 476, 73

\bibitem[{Girardi {et~al.}(2000)Girardi, Bressan, Bertelli, \&
  Chiosi}]{Girardi2000}
Girardi, L., Bressan, A., Bertelli, G., \& Chiosi, C. 2000, A\&AS, 141, 371

\bibitem[{Girardi {et~al.}(2008)Girardi, Dalcanton, Williams, Jong, Gallart,
  Monelli, Groenewegen, Holtzman, Olsen, Seth, Weisz, \&
  collaboration)}]{Girardi2008}
Girardi, L., Dalcanton, J., Williams, B., {et~al.} 2008, PASP, 120, 583

\bibitem[{Girardi {et~al.}(2010)Girardi, Williams, Gilbert, Rosenfield,
  Dalcanton, Marigo, Boyer, Dolphin, Weisz, Melbourne, Olsen, Seth, \&
  Skillman}]{Girardi2010}
Girardi, L., Williams, B.~F., Gilbert, K.~M., {et~al.} 2010, ApJ, 724, 1030

\bibitem[{Groth(1986)}]{Groth1986}
Groth, E.~J. 1986, AJ, 91, 1244

\bibitem[{Hunter(2007)}]{Hunter:2007}
Hunter, J.~D. 2007, Computing In Science \& Engineering, 9, 90

\bibitem[{{Jennings} {et~al.}(2012){Jennings}, {Williams}, {Murphy},
  {Dalcanton}, {Gilbert}, {Dolphin}, {Fouesneau}, \& {Weisz}}]{Jennings2012}
{Jennings}, Z.~G., {Williams}, B.~F., {Murphy}, J.~W., {et~al.} 2012, \apj,
  761, 26

\bibitem[{{Johnson} {et~al.}(2015){Johnson}, {Seth}, {Dalcanton}, {Wallace},
  {Simpson}, {Lintott}, {Kapadia}, {Skillman}, {Caldwell}, {Fouesneau},
  {Weisz}, {Williams}, {Beerman}, {Gouliermis}, \& {Sarajedini}}]{Johnson2015}
{Johnson}, L.~C., {Seth}, A.~C., {Dalcanton}, J.~J., {et~al.} 2015, \apj, 802,
  127

\bibitem[{Kippenhahn \& Smith(1969)}]{Kippenhahn1969}
Kippenhahn, R., \& Smith, L. 1969, A\&A, 1, 142

\bibitem[{Kodric {et~al.}(2013)Kodric, Riffeser, Hopp, Seitz, Koppenhoefer,
  Bender, Goessl, Snigula, Lee, Ngeow, Chambers, Magnier, Price, Burgett,
  Hodapp, Kaiser, \& Kudritzki}]{Kodric2013}
Kodric, M., Riffeser, A., Hopp, U., {et~al.} 2013, AJ, 145, 106

\bibitem[{{Kodric} {et~al.}(2015){Kodric}, {Riffeser}, {Seitz}, {Snigula},
  {Hopp}, {Lee}, {Goessl}, {Koppenhoefer}, {Bender}, \& {Gieren}}]{Kodric2015}
{Kodric}, M., {Riffeser}, A., {Seitz}, S., {et~al.} 2015, \apj, 799, 144

\bibitem[{{Leavitt} \& {Pickering}(1912)}]{Leavitt1912}
{Leavitt}, H.~S., \& {Pickering}, E.~C. 1912, Harvard College Observatory
  Circular, 173, 1

\bibitem[{Lee {et~al.}(2012)Lee, Riffeser, Koppenhoefer, Seitz, Bender, Hopp,
  Gössl, Saglia, Snigula, Sweeney, Burgett, Chambers, Grav, Heasley, Hodapp,
  Kaiser, Magnier, Morgan, Price, Stubbs, Tonry, \& Wainscoat}]{Lee2012}
Lee, C.-H., Riffeser, A., Koppenhoefer, J., {et~al.} 2012, AJ, 143, 89

\bibitem[{Macri(2004)}]{Macri2004}
Macri, L.~M. 2004, in , eprint: {arXiv:astro-ph/0310016}, 33

\bibitem[{Magnier {et~al.}(1997)Magnier, Prins, Augusteijn, van Paradijs, \&
  Lewin}]{Magnier1997}
Magnier, E.~A., Prins, S., Augusteijn, T., van Paradijs, J., \& Lewin, W. H.~G.
  1997, A\&A, 326, 442

\bibitem[{Marigo {et~al.}(2008)Marigo, Girardi, Bressan, Groenewegen, Silva, \&
  Granato}]{Marigo2008}
Marigo, P., Girardi, L., Bressan, A., {et~al.} 2008, A\&A, 482, 883

\bibitem[{Mochejska {et~al.}(2000)Mochejska, Macri, Sasselov, \&
  Stanek}]{Mochejska2000}
Mochejska, B.~J., Macri, L.~M., Sasselov, D.~D., \& Stanek, K.~Z. 2000, AJ,
  120, 810

\bibitem[{P\'erez \& Granger(2007)}]{PER-GRA:2007}
P\'erez, F., \& Granger, B.~E. 2007, Computing in Science and Engineering, 9,
  21

\bibitem[{Pietrinferni {et~al.}(2004)Pietrinferni, Cassisi, Salaris, \&
  Castelli}]{Pietrinferni2004}
Pietrinferni, A., Cassisi, S., Salaris, M., \& Castelli, F. 2004, ApJ, 612, 168

\bibitem[{Pietrzynski \& Udalski(1999)}]{Pietrzynski1999}
Pietrzynski, G., \& Udalski, A. 1999, Acta Astron., 49, 543

\bibitem[{Riess {et~al.}(2012)Riess, Fliri, \& Valls-Gabaud}]{Riess2012}
Riess, A.~G., Fliri, J., \& Valls-Gabaud, D. 2012, ApJ, 745, 156

\bibitem[{Schaller {et~al.}(1992)Schaller, Schaerer, Meynet, \&
  Maeder}]{Schaller1992}
Schaller, G., Schaerer, D., Meynet, G., \& Maeder, A. 1992, A\&AS, 96, 269

\bibitem[{Sirianni {et~al.}(2005)Sirianni, Jee, Benítez, Blakeslee, Martel,
  Meurer, Clampin, Marchi, Ford, Gilliland, Hartig, Illingworth, Mack, \&
  {McCann}}]{Sirianni2005}
Sirianni, M., Jee, M., Benítez, N., {et~al.} 2005, PASP, 117, 1049

\bibitem[{Tammann {et~al.}(2003)Tammann, Sandage, \& Reindl}]{Tammann2003}
Tammann, G.~A., Sandage, A., \& Reindl, B. 2003, A\&A, 404, 423

\bibitem[{Tonry {et~al.}(2012)Tonry, Stubbs, Lykke, Doherty, Shivvers, Burgett,
  Chambers, Hodapp, Kaiser, Kudritzki, Magnier, Morgan, Price, \&
  Wainscoat}]{Tonry2012}
Tonry, J.~L., Stubbs, C.~W., Lykke, K.~R., {et~al.} 2012, ApJ, 750, 99

\bibitem[{Vilardell {et~al.}(2007)Vilardell, Jordi, \& Ribas}]{Vilardell2007}
Vilardell, F., Jordi, C., \& Ribas, I. 2007, A\&A, 473, 847

\bibitem[{{Wagner-Kaiser} {et~al.}(2015){Wagner-Kaiser}, {Sarajedini},
  {Dalcanton}, {Williams}, \& {Dolphin}}]{WagnerKaiser2015}
{Wagner-Kaiser}, R., {Sarajedini}, A., {Dalcanton}, J.~J., {Williams}, B.~F.,
  \& {Dolphin}, A. 2015, ArXiv e-prints, arXiv:1504.05118

\bibitem[{Welch \& Stetson(1993)}]{Welch1993}
Welch, D.~L., \& Stetson, P.~B. 1993, AJ, 105, 1813

\bibitem[{{Williams} {et~al.}(2014){Williams}, {Lang}, {Dalcanton}, {Dolphin},
  {Weisz}, {Bell}, {Bianchi}, {Byler}, {Gilbert}, {Girardi}, {Gordon},
  {Gregersen}, {Johnson}, {Kalirai}, {Lauer}, {Monachesi}, {Rosenfield},
  {Seth}, \& {Skillman}}]{Williams2014}
{Williams}, B.~F., {Lang}, D., {Dalcanton}, J.~J., {et~al.} 2014, \apjs, 215, 9

\end{thebibliography}
\bibliographystyle{apj}

\newpage

\appendix

Here we present the diagnostic plots produced during Cepheid identification and dephasing described in Section~\ref{analysis}.
Refer to the caption of Figure~\ref{fig:cmdvar} for a description of the various panels and symbols.
In each case, the putative Cepheid is identified quantitatively by the variability metrics (bottom panels); and its HST measurements are compared to the PS1 data (top panel).

\begin{figure}
    \centering
    \includegraphics[scale=0.6]{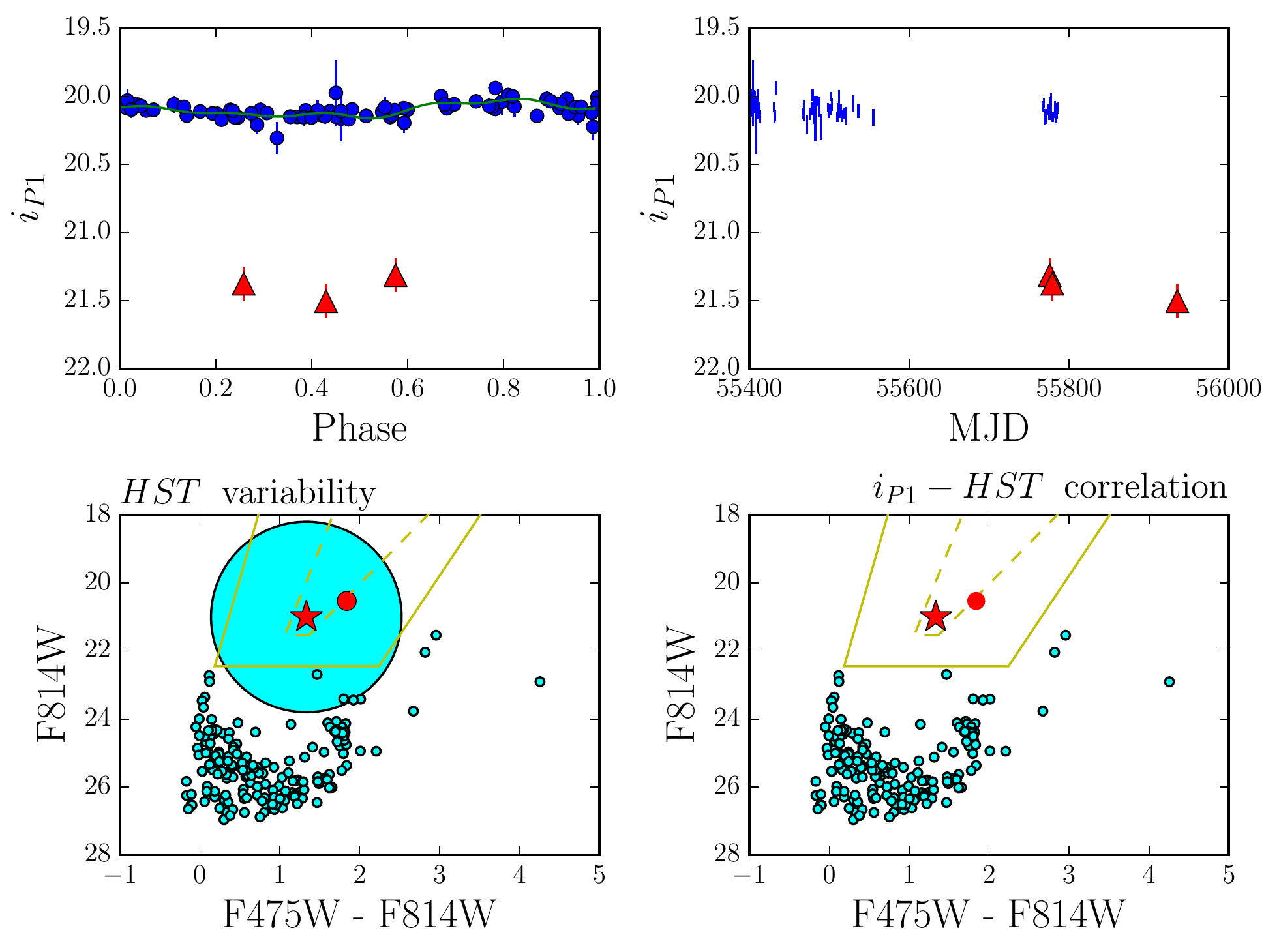}
    \caption{Cepheid identification for CC1; see Figure~\ref{fig:cmdvar} for explanation of plot symbols.}
\end{figure}

\begin{figure}
    \centering
    \includegraphics[scale=0.6]{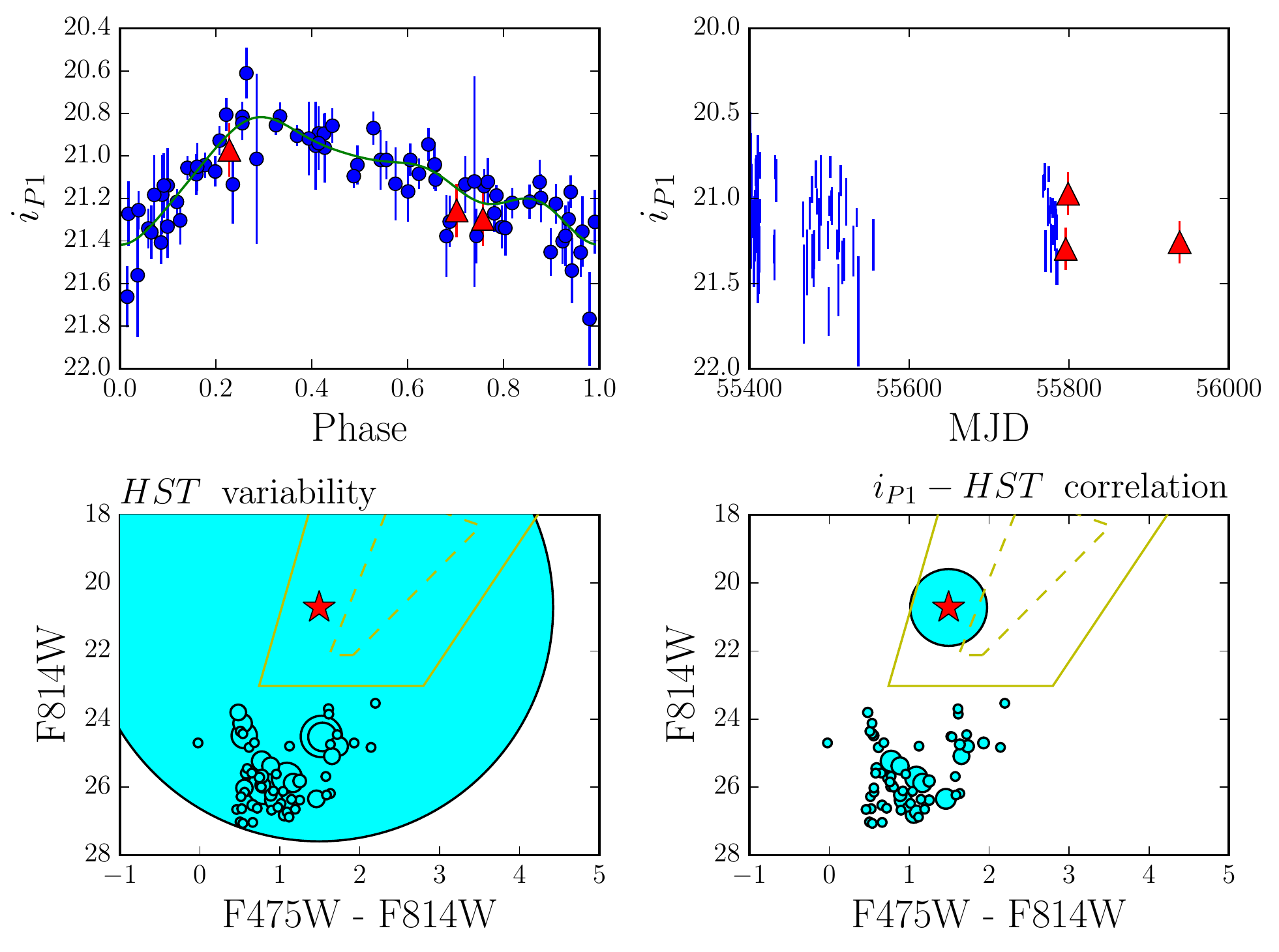}
    \caption{Cepheid identification for CC3; see Figure~\ref{fig:cmdvar} for explanation of plot symbols.}
\end{figure}

\begin{figure}
    \centering
    \includegraphics[scale=0.6]{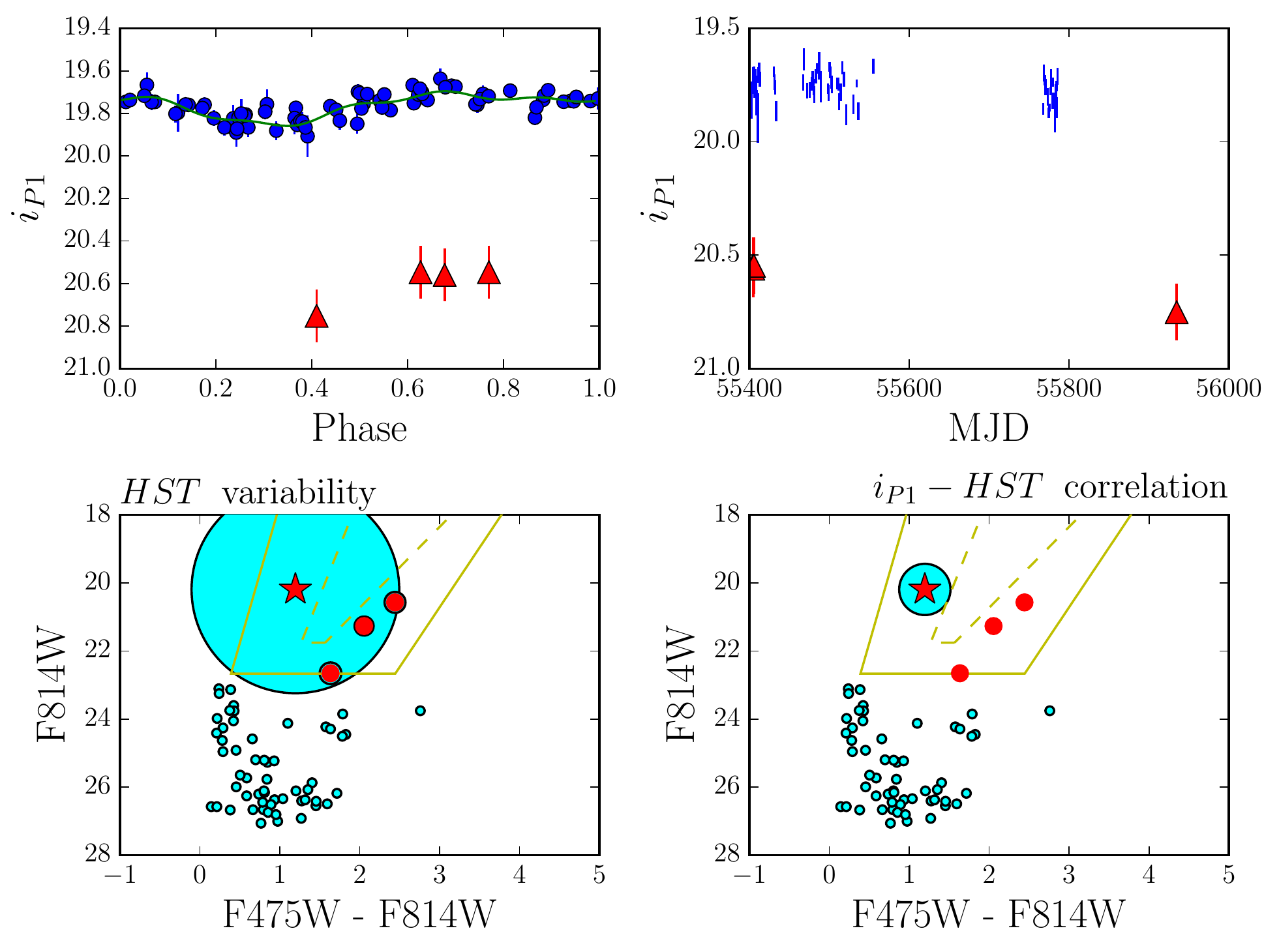}
    \caption{Cepheid identification for CC4; see Figure~\ref{fig:cmdvar} for explanation of plot symbols.}
\end{figure}

\begin{figure}
    \centering
    \includegraphics[scale=0.6]{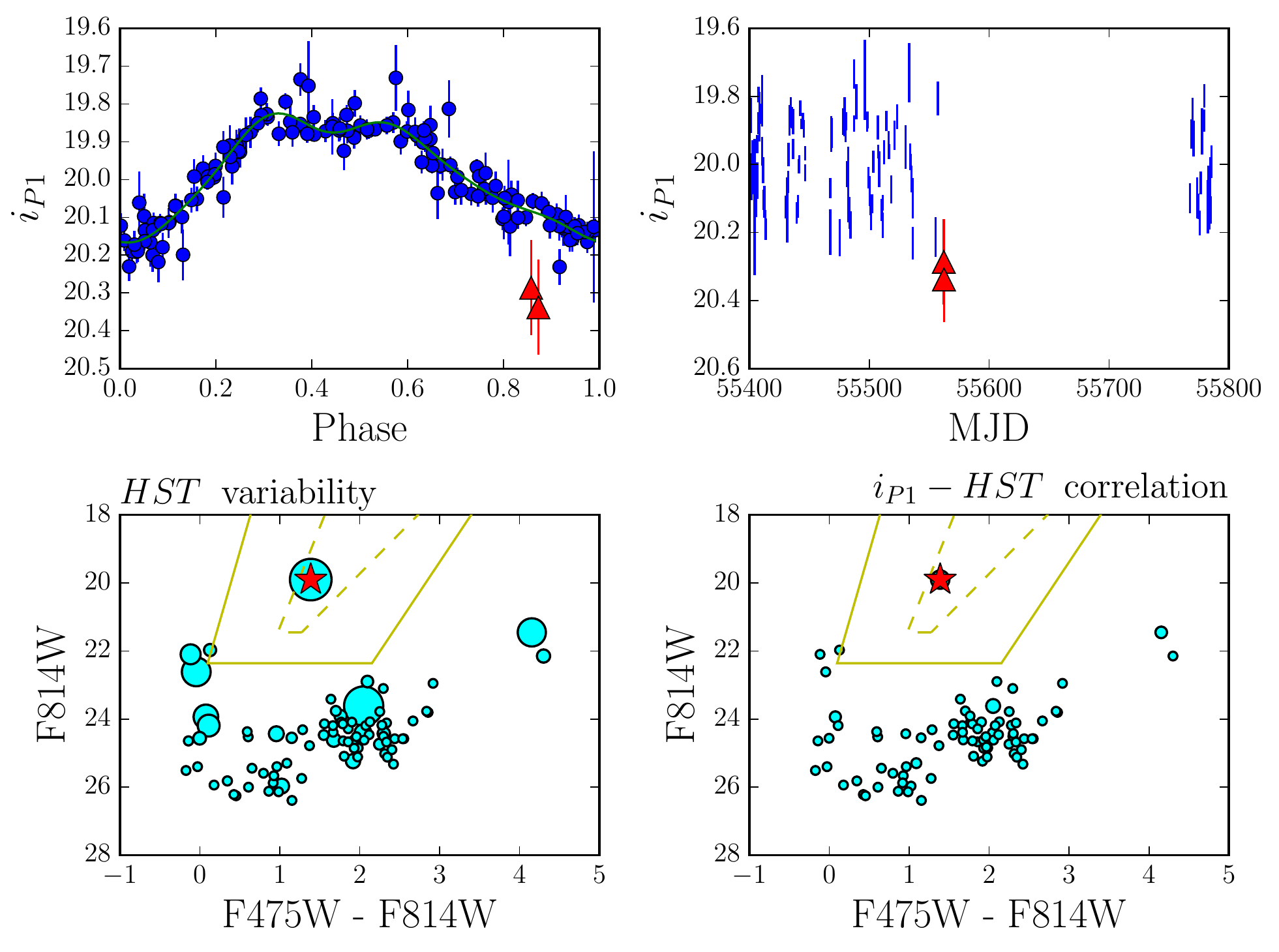}
    \caption{Cepheid identification for CC5; see Figure~\ref{fig:cmdvar} for explanation of plot symbols.}
\end{figure}

\begin{figure}
    \centering
    \includegraphics[scale=0.6]{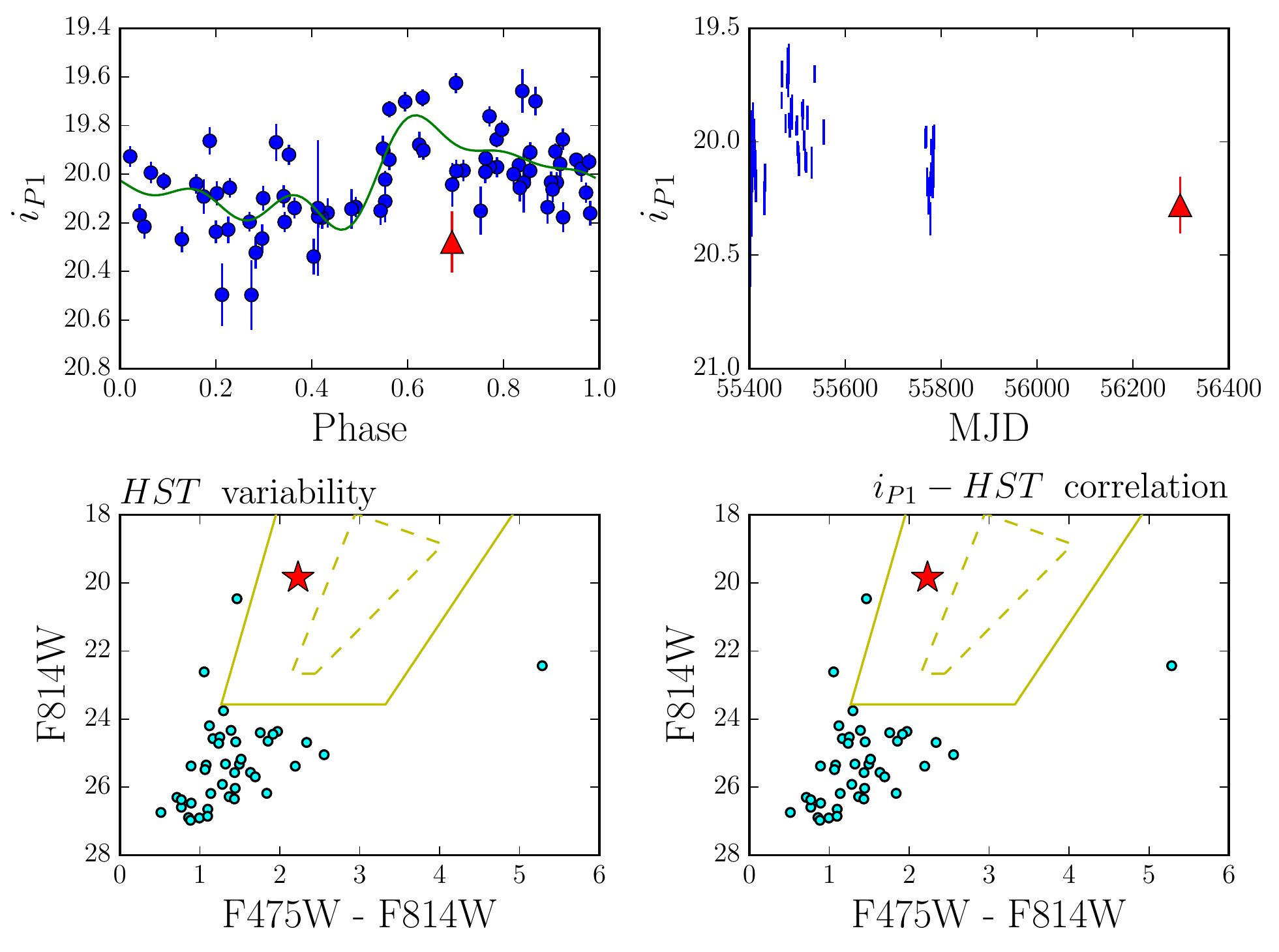}
    \caption{Cepheid identification for CC6; see Figure~\ref{fig:cmdvar} for explanation of plot symbols. Note that this cluster was only observed in one PHAT field, and thus no variability information is available for the HST CMDs. Accordingly, the point size in the bottom panels is fixed.}
\end{figure}

\begin{figure}
    \centering
    \includegraphics[scale=0.6]{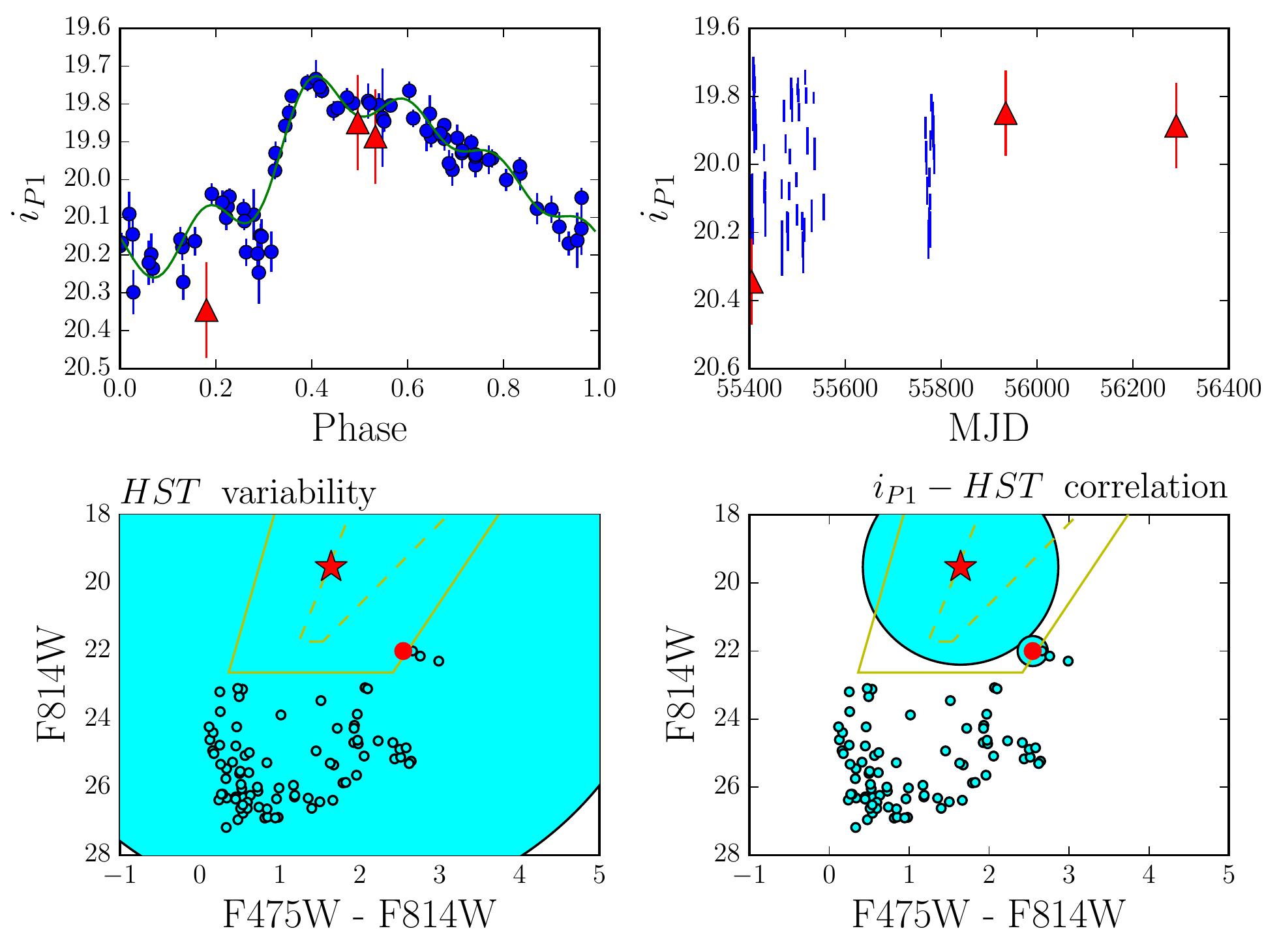}
    \caption{Cepheid identification for CC7; see Figure~\ref{fig:cmdvar} for explanation of plot symbols.}
\end{figure}

\begin{figure}
    \centering
    \includegraphics[scale=0.6]{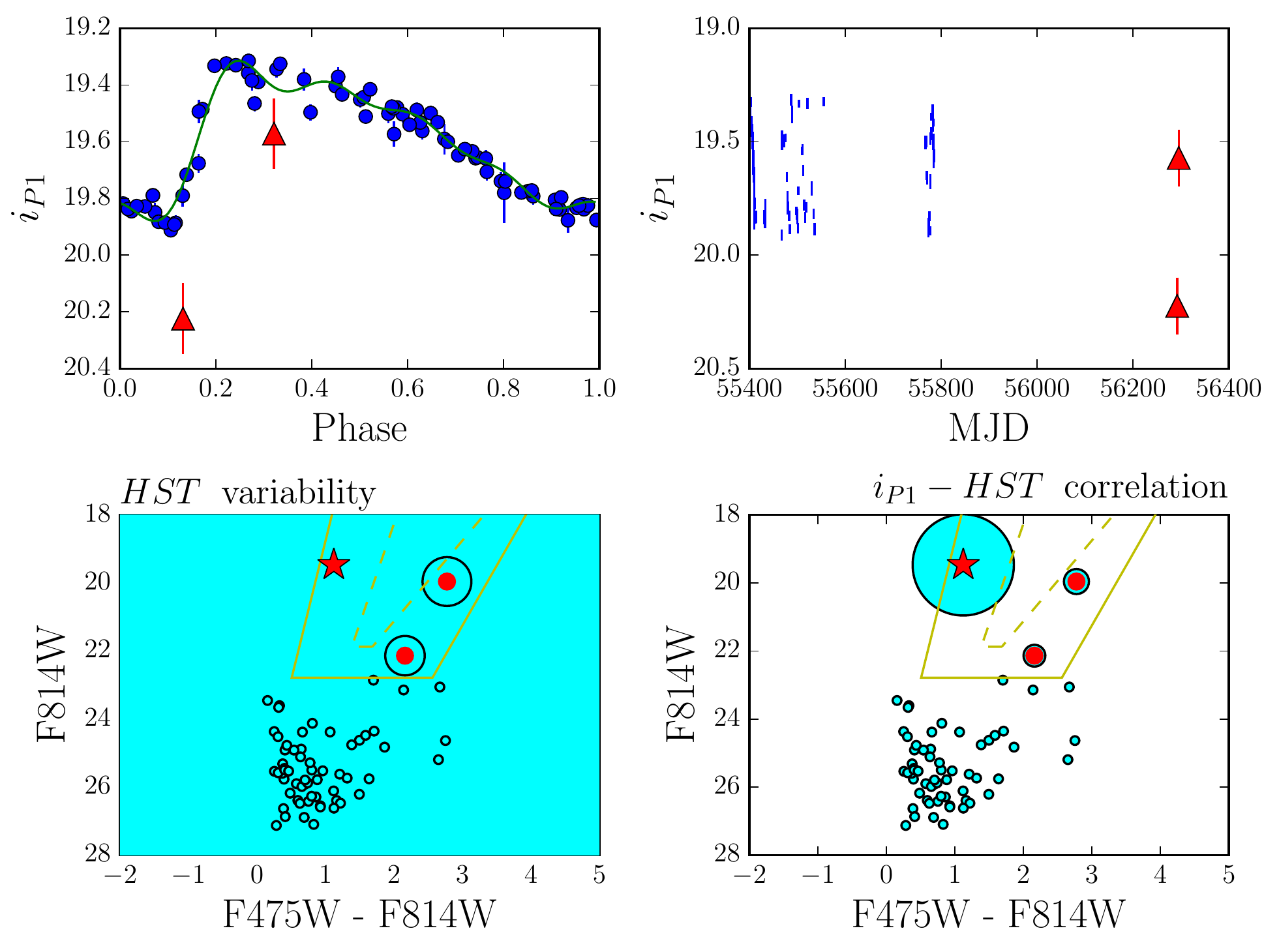}
    \caption{Cepheid identification for CC8; see Figure~\ref{fig:cmdvar} for explanation of plot symbols.}
\end{figure}

\begin{figure}
    \centering
    \includegraphics[scale=0.6]{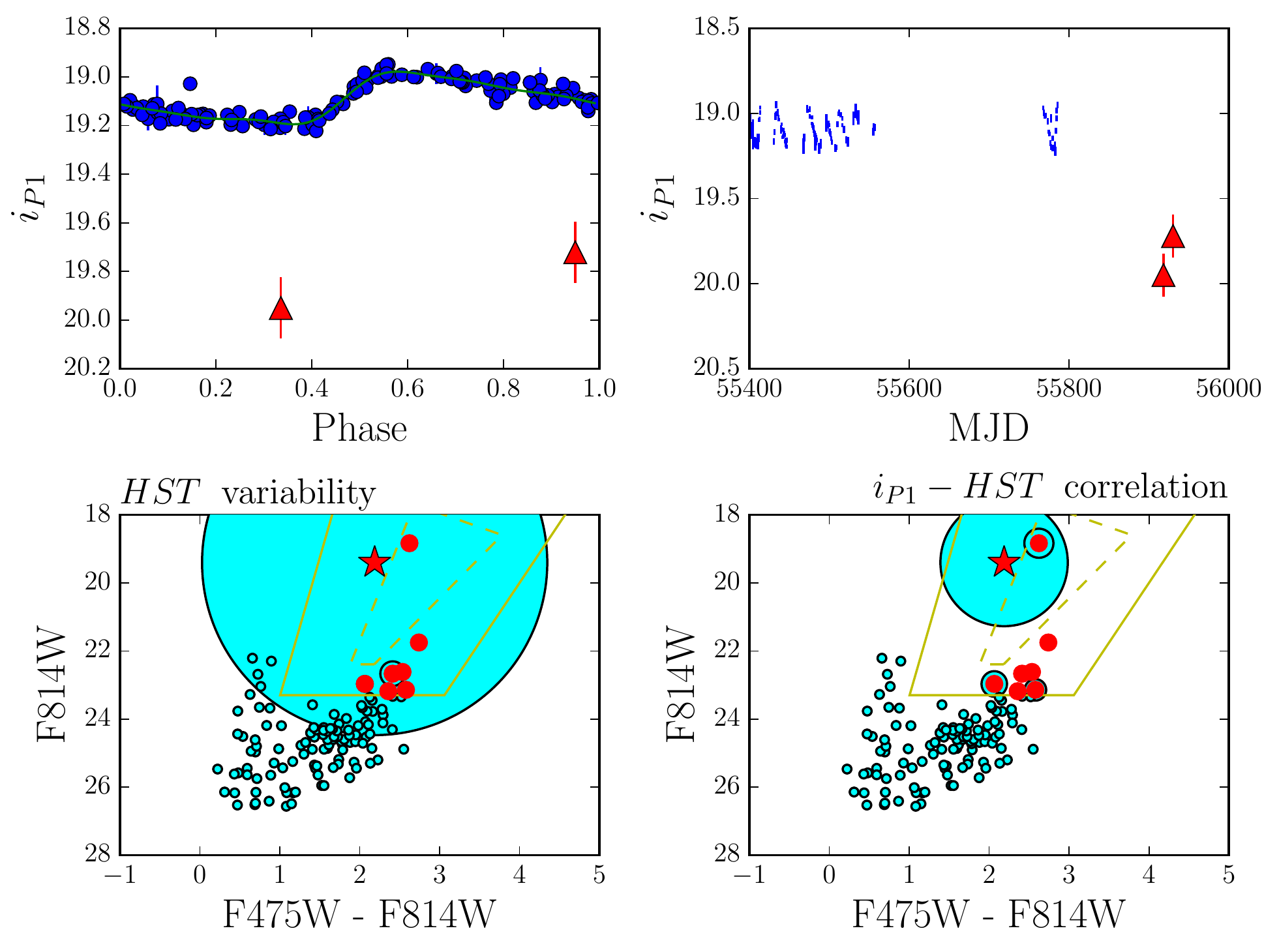}
    \caption{Cepheid identification for CC9; see Figure~\ref{fig:cmdvar} for explanation of plot symbols.}
\end{figure}

\begin{figure}
    \centering
    \includegraphics[scale=0.6]{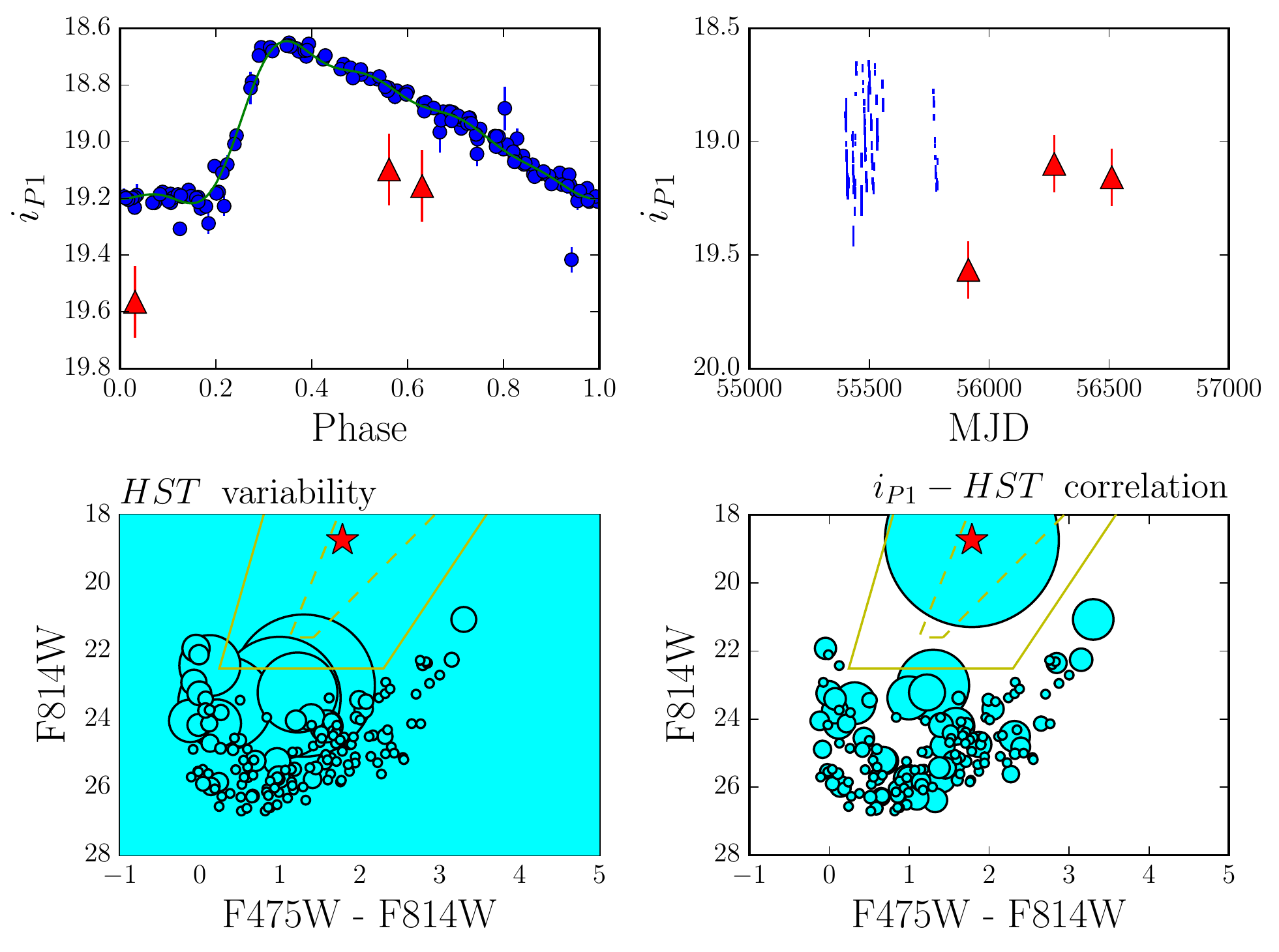}
    \caption{Cepheid identification for CC10; see Figure~\ref{fig:cmdvar} for explanation of plot symbols.}
\end{figure}

\begin{figure}
    \centering
    \includegraphics[scale=0.6]{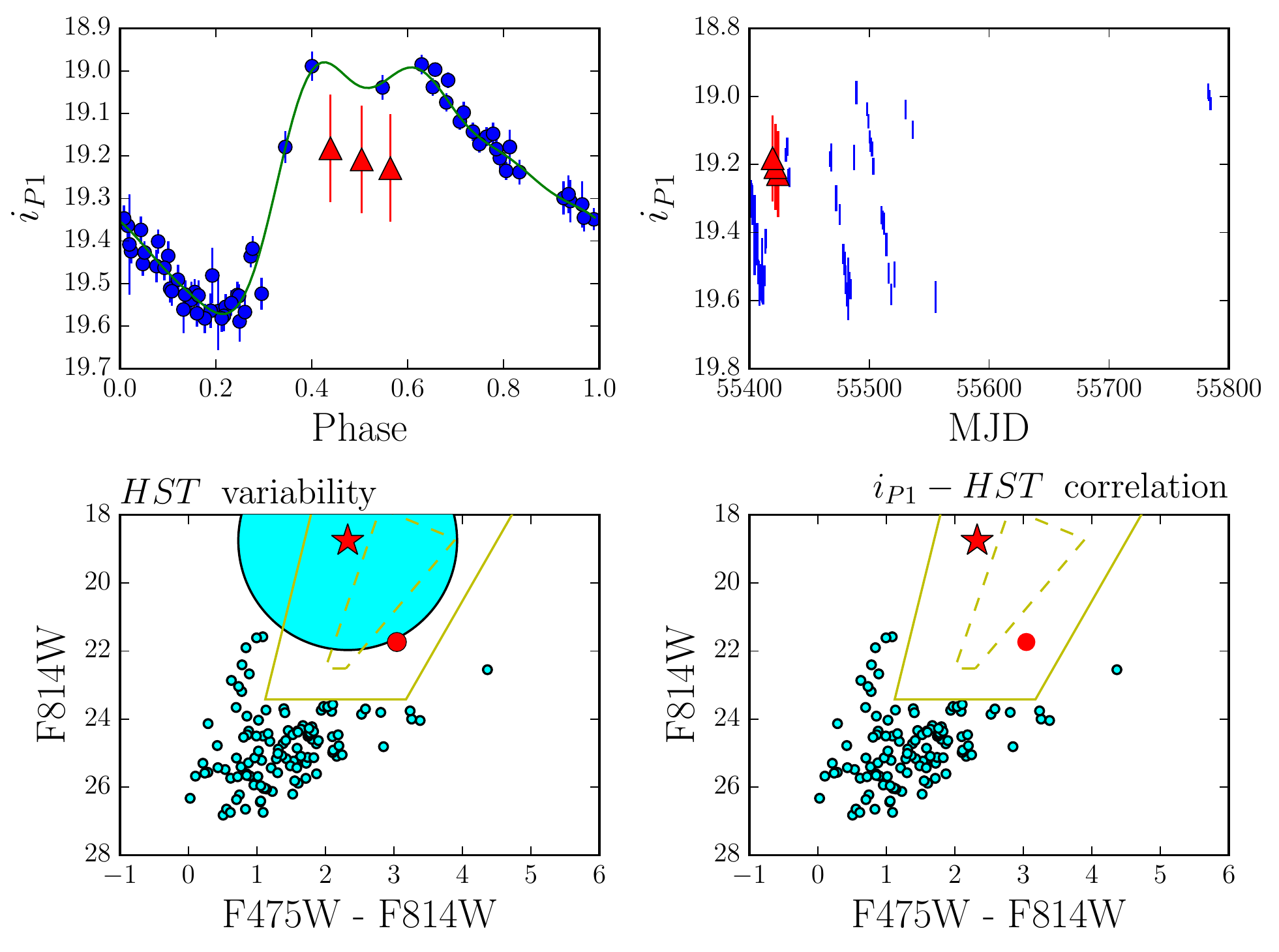}
    \caption{Cepheid identification for CC11; see Figure~\ref{fig:cmdvar} for explanation of plot symbols.}
\end{figure}
\end{document}